\documentclass[a4paper,11pt]{article}
\pdfoutput=1 % if your are submitting a pdflatex (i.e. if you have
\usepackage{jcappub} % for details on the use of the package, please
\usepackage[T1]{fontenc} % if needed

\title{The PAU Survey: Measurement of Narrow-band galaxy properties with Approximate Bayesian Computation}

\author[a]{Luca Tortorelli,}
\author[b,c]{Malgorzata Siudek,}
\author[a]{Beatrice Moser,}
\author[a]{Tomasz Kacprzak,}
\author[a]{Pascale Berner,}
\author[a]{Alexandre Refregier,}
\author[d]{Adam Amara,}
\author[e]{Juan Garc\'ia-Bellido,}
\author[b]{Laura Cabayol,}
\author[b,1]{Jorge Carretero,}
\author[f,g]{Francisco J. Castander,}
\author[h]{Juan De Vicente,}
\author[b]{Martin Eriksen,}
\author[b]{Enrique Fernandez,}
\author[f,g]{Enrique Gaztanaga,}
\author[i]{Hendrik Hildebrandt,}
\author[j]{Benjamin Joachimi,}
\author[b,k]{Ramon Miquel,}
\author[h]{Ignacio Sevilla-Noarbe,}
\author[b]{Crist\'obal Padilla,}
\author[f,g]{Pablo Renard,}
\author[h]{Eusebio Sanchez,}
\author[f,g]{Santiago Serrano,}
\author[h,1]{Pau Tallada-Cresp\'i\note{also at Port d'Informaci\'o Cient\'ifica (PIC), Campus UAB, c/ de l'Albareda, Edifici D, 08193 Barcelona, Spain}}
\author[i]{and Angus H. Wright}

\affiliation[a]{Institute for Particle Physics and Astrophysics, ETH Z\"urich, Wolfgang-Pauli-Str. 27, 8093 Z\"urich, Switzerland}
\affiliation[b]{Institut de F\'isica d'Altes Energies (IFAE), The Barcelona Institute of Science and Technology, 08193 Bellaterra (Barcelona), Spain}
\affiliation[c]{National Centre for Nuclear Research, ul. Pasteura 7, 02-093 Warsaw, Poland}
\affiliation[d]{Institute of Cosmology and Gravitation, University of Portsmouth, Dennis Sciama Building, Burnaby Road, Portsmouth PO1 3FX, U.K.}
\affiliation[e]{Instituto de F\'isica Te\'orica (UAM/CSIC), Cantoblanco, E-28049 Madrid, Spain}
\affiliation[f]{Institute of Space Sciences (ICE, CSIC), Carrer de Can Magrans s/n, E-08193 Bellaterra (Barcelona), Spain}
\affiliation[g]{Institut d'Estudis Espacials de Catalunya (IEEC), E-08034 Barcelona, Spain}
\affiliation[h]{Centro de Investigaciones Energ\'eticas, Medioambientales y Tecnol\'ogicas (CIEMAT), Avenida Complutense 40, E-28040 Madrid, Spain}
\affiliation[i]{Ruhr University Bochum, Faculty of Physics and Astronomy, Astronomical Institute (AIRUB), German Centre for Cosmological Lensing, 44780 Bochum, Germany}
\affiliation[j]{Department of Physics and Astronomy, University College London, Gower Street, London WC1E 6BT, UK}
\affiliation[k]{Instituci\'o Catalana de Recerca i Estudis Avan\c{c}ats (ICREA), 08010 Barcelona, Spain}

\emailAdd{torluca@phys.ethz.ch}

\abstract{Narrow-band imaging surveys allow the study of the spectral characteristics of galaxies without the need of performing their spectroscopic follow-up. In this work, we forward-model the Physics of the Accelerating Universe Survey (PAUS) narrow-band data. The aim is to improve the constraints on the spectral coefficients used to create the galaxy spectral energy distributions (SED) of the galaxy population model in Tortorelli et al. 2020. In that work, the model parameters were inferred from the Canada-France-Hawaii Telescope Legacy Survey (CFHTLS) data using Approximate Bayesian Computation (ABC). This led to stringent constraints on the B-band galaxy luminosity function parameters, but left the spectral coefficients only broadly constrained. To address that, we perform an ABC inference using CFHTLS and PAUS data. This is the first time our approach combining forward-modelling and ABC is applied simultaneously to multiple datasets. We test the results of the ABC inference by comparing the narrow-band magnitudes of the observed and simulated galaxies using Principal Component Analysis, finding a very good agreement. Furthermore, we prove the scientific potential of the constrained galaxy population model to provide realistic stellar population properties by measuring them with the SED fitting code \textsc{CIGALE}. We use CFHTLS broad-band and PAUS narrow-band photometry for a flux-limited ($\mathrm{i}<22.5$) sample of galaxies up to redshift $\mathrm{z \sim 0.8}$. We find that properties like stellar masses, star-formation rates, mass-weighted stellar ages and metallicities are in agreement within errors between observations and simulations. Overall, this work shows the ability of our galaxy population model to correctly forward-model a complex dataset such as PAUS and the ability to reproduce the diversity of galaxy properties at the redshift range spanned by CFHTLS and PAUS.}

\begin{document}
\maketitle
\flushbottom

\section{Introduction}
\label{section:introduction}

Spectroscopic and photometric broad-band galaxy surveys are complementary tools to probe the physical properties, the matter content and the geometry of our Universe. Spectroscopic surveys allow for precise redshift estimation and detailed physical properties extraction of galaxies. However, most of the spectroscopic surveys are flux-limited at brighter magnitudes than broad-band surveys. This is mainly due to spectroscopy being time-consuming and therefore requiring large telescopes and long integration times. Broad-band photometric surveys allow us to probe large volumes of the Universe, significantly reducing the impact of cosmic variance, but at the cost of lacking precise redshift information to allow for a tomographic analysis of galaxy properties. 

With the growing number of wide-field surveys and therefore galaxies to be analysed, new methods to measure galaxy population properties need to be developed. In \cite{Tortorelli2020}, we proposed a new method to measure the luminosity function (hereafter, LF) that relies on the forward-modelling of photometric and spectroscopic wide-field surveys and on Approximate Bayesian Computation (hereafter, ABC) \cite{akeret15}. We constrained the empirical galaxy population model developed in \cite{herbel17} using the Canada-France-Hawaii Telescope Legacy Survey (hereafter, CFHTLS) \cite{cuillandre12} broad-band images. This allowed us to obtain the first measurement in literature of galaxy population properties, specifically the B-band galaxy LF, using ABC.

CFHTLS data provided strong constraints on the LF and on the size distribution of galaxies at different redshifts, thanks to their depth and sky coverage. However, being a broad-band survey, it lacks in spectral resolution. This implies that the coefficients used to assign spectral energy distributions to galaxies (hereafter, spectral coefficients) in the galaxy population model of \cite{Tortorelli2020} remained only broadly constrained.

Narrow-band data are the perfect observational dataset to improve upon the results in \cite{Tortorelli2020} and constrain the spectral coefficients of the galaxy population model. Narrow-band galaxy surveys offer several advantages with respect to both spectroscopic and photometric broad-band galaxy surveys. They allow us to probe large portions of the sky in a single observation with good spectral resolution. They are more efficient in observing faint galaxies compared to spectroscopic instruments. They do not require a target selection, unlike  spectroscopic surveys, since all objects brighter than the limiting magnitude of the survey, if non-detections are counted, get a narrow-band spectrum.

Narrow-bands have the the advantage of allowing the measurement of accurate photometric redshifts ($\sigma_{\mathrm{z}} \sim 0.0035$, \cite{Eriksen2019,Alarcon2021}), but they have been mainly used so far to select and measure the fluxes of emission lines (e.g., H${\alpha}$ narrow-band surveys \cite{James2004,Drew2005,Kennicutt2008}, narrow-band selected [OIII] emission line galaxies \cite{Ly2012,Suzuki2015}). It is only recently, starting from COMBO-17 and COSMOS-30 \cite{Wolf2003,Bell2004,Ilbert2009}, that extensive narrow-band and medium-band photometric galaxy surveys have taken place \cite{Moles2008,Benitez2014,Cenarro2019}. The Physics of the Accelerating Universe Survey (hereafter, PAUS) \cite{Benitez2009,marti14} is one of those. PAUS is a 40 narrow-band optical survey aimed at providing very high photo-z precision and at measuring galaxy properties from narrow-band fluxes. 

There is a strong interest in measuring galaxy properties with narrow-bands. Despite being low resolution, the narrow-band spectrum contains enough information to derive galaxy properties such as stellar masses, star-formation rates (hereafter, SFRs), ages and more. Measuring galaxy properties with narrow-bands is a young field where very few attempts have been carried out so far (e.g., MUFFIT \cite{Diaz-Garcia2015,SanRoman2018,Gonzalez-Delgado2021,Liew-Cain2021}) and where many widely used SED fitting codes (e.g., \textsc{CIGALE} \cite{Burgarella2005,Noll2009,Boquien2019}, \textsc{Prospector} \cite{Johnson2021}) have not been tested yet.

In this work, we use broad-band data from CFHTLS and narrow-band data from PAUS. This is the first time our approach combining forward-modelling and ABC is applied simultaneously to multiple datasets. The combination of the two datasets allows for all aspects of the galaxy population model in \cite{Tortorelli2020} to be constrained, providing a fully data-calibrated empirical galaxy population model. To constrain it, we use the likelihood-free inference method ABC. Many examples of exoplanets, galaxy evolution  and cosmological results with likelihood-free methods exist nowadays in the literature \cite{Cameron2012,Weyant2013,Jennings2016,Hahn2017,Alsing2018a,Alsing2018b,Witzel2018,Hsu2018,Kacprzak2018,Leclercq2019,Alsing2019a,Alsing2019b,Aufort2020,Plazas2020,Zunli2020,Kunimoto2020,Kacprzak2020,Lemos2021,Livet2021,Jeffrey2021,Weiss2021}. Our combined approach of ABC and forward-modelling allows us to create a full end-to-end simulation of galaxy images and spectra, by including all the relevant observational effects. This is fundamental since, in order for the forward-modelling approach to be successful, the exact same analysis steps and selection cuts need to be performed on observed data and simulations. This is possible thanks to a realistic and fast image simulator, namely the Ultra Fast image generator (hereafter, UFig) \cite{berge13,bruderer16}.

After obtaining the approximate posterior distribution of the galaxy population model parameters from the ABC inference, we draw simulated galaxies from the model and we measure their stellar population properties with SED fitting. We do the same on observed data, using both broad-band and narrow-band magnitudes. This allows us to test how well the galaxy population model is able to predict galaxy stellar properties that are not directly drawn from the model itself. Furthermore, it allows us to also test the performance of the SED fitting code \textsc{CIGALE} to recover the stellar population properties of observed and simulated galaxies with narrow-bands.

The paper is structured as follows. Section \ref{section:data} gives a description of the data that we use in this work. Section \ref{section:gal_pop_model} provides a brief description of the image simulator and of the galaxy population model. Section \ref{section:method} describes the method we develop in this work. Section \ref{section:abc_cfhtls_paus} describes the ABC inference we perform on CFHTLS and PAUS data. Using galaxies from the posterior distribution, we perform a Principal Component Analysis in section \ref{section:pca_nb_spectra} and we measure stellar population properties with SED fitting in section \ref{section:stellar_pop}. We draw our conclusions and provide future directions in section \ref{section:conclusions}. Throughout this work, we use a standard $\mathrm{\Lambda CDM}$ cosmology with $\Omega_{\mathrm{m}} = 0.3$, $\Omega_{\Lambda} = 0.7$ and $\mathrm{H_0 = 70\ km\ s^{-1}\ Mpc^{-1} }$. 

\section{Data}
\label{section:data}

This section describes the data we use to constrain the galaxy population model calibrated in \cite{Tortorelli2020} and, in particular, its spectral coefficients. We use broad-band data from CFHTLS and narrow-band data from PAUS. Narrow-band photometry is noisier than broad-band photometry. The narrowness of the waveband probed reduces the amount of flux collected from a source, therefore making its detection more challenging. For this reason, we use broad-band images to detect sources.

\subsection{Canada-France-Hawaii Telescope Legacy Survey}
\label{subsection:data_cfhtls}

\begin{figure}[t!]
\centering
\includegraphics[width=15.5cm]{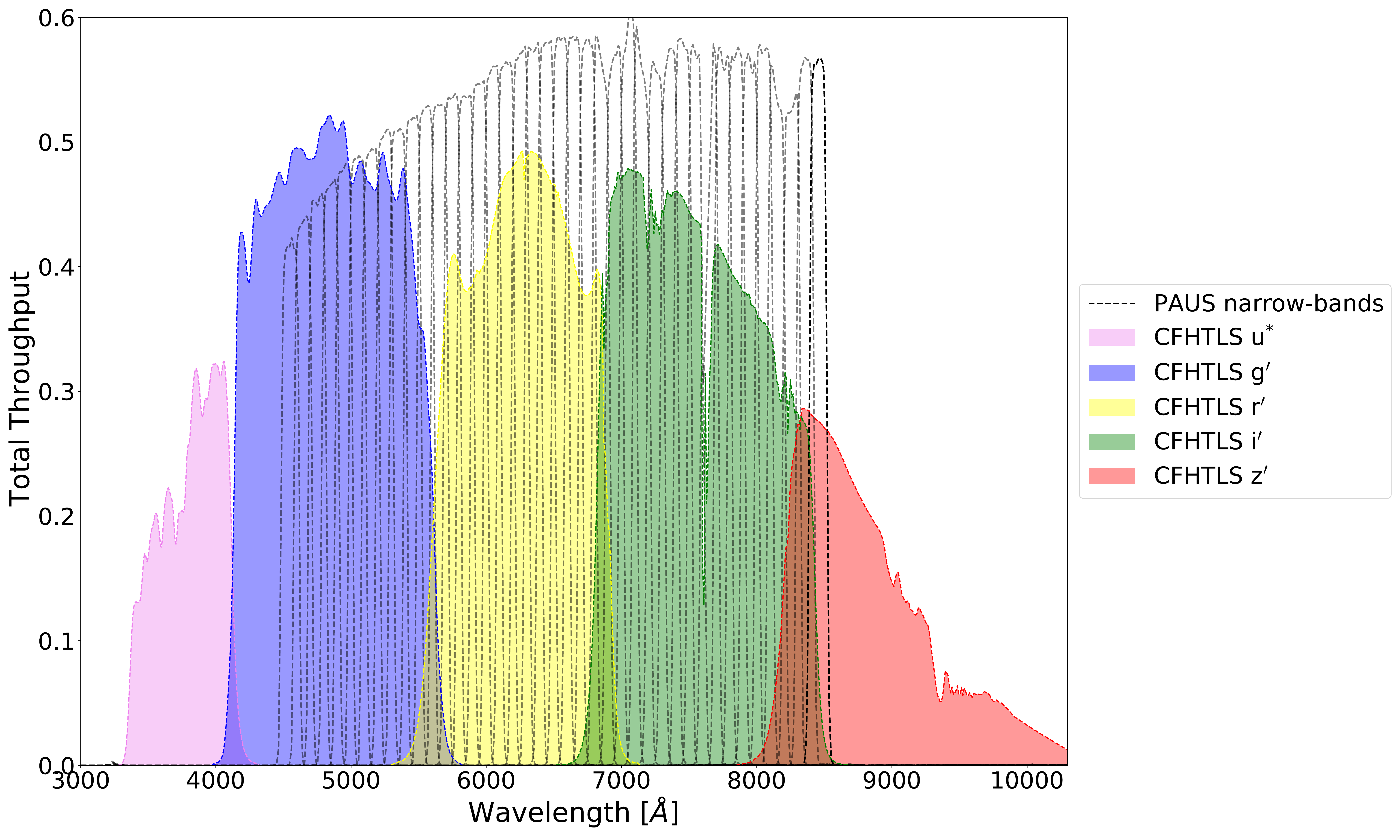}
\caption{This figure shows the total throughput and wavelength coverage of CFHTLS broad-band (colour-filled areas) and PAUS narrow-band filters (black dashed lines). CFHTLS filters span the wavelength range $3300 - 10600$ \AA, while PAUS narrow-bands span the wavelength range $4500-8500$ \AA.} 
\label{fig:tortorelli_fig1}
\end{figure}

CFHTLS is a wide-field photometric galaxy survey conducted with the CFHT telescope on top of Mauna Kea (Hawaii, USA), using the MegaCam camera \cite{boulade00}. It covers $157$ square degrees in five optical bands $\mathrm{u^{*}}$, $\mathrm{g'}$, $\mathrm{r'}$, $\mathrm{i'}$, $\mathrm{z'}$ (see figure \ref{fig:tortorelli_fig1}) with a median sub-arcsecond seeing. The survey is conducted with two different levels of image depth. CFHTLS `Wide' is a $155$ square degrees survey in four independent contiguous patches (W1', `W2', `W3', `W4'). It has a $80\%$ completeness limit in AB of $\mathrm{u^{*}}=25.2$, $\mathrm{g'}=25.5$, $\mathrm{r'}=25.0$, $\mathrm{i'}=24.8$, $\mathrm{z'}=23.9$ for point sources. CFHTLS `Deep' consists of 4 ultra deep pointings (`D1', `D2', `D3', `D4'). It has a $80\%$ completeness limit in AB of $\mathrm{u^{*}}=26.3$, $\mathrm{g'}=26.0$, $\mathrm{r'}=25.6$, $\mathrm{i'}=25.4$, $\mathrm{z'}=25.0$ for point sources. We use data from the 7th and final data release of CFHTLS \cite{cuillandre12}. Each CFHTLS image covers one square degree with a pixel scale of $0.186$ arcsec. Among the main scientific themes that shaped the survey strategy, there were the search of Type Ia supernovae, the cosmic shear analysis and the census of Kuiper belt objects. In our analysis, we use the `Wide' fields due to the large sky coverage and to the faster simulation time with respect to the `Deep' images, given their different depths. In order to mitigate the effect of cosmic variance, we use widely-separated fields, as in this case the fluctuations in each field will be independent. Using larger fields also help to mitigate cosmic variance, although at a lesser degree since the large-scale modes in the matter power spectrum will couple density fluctuations across a field. Sampling more, smaller, widely-separated fields is in general the optimal strategy for minimising sample/cosmic variance at fixed total sky area \cite{Newman2015}. The large CFHTLS `Wide' sky coverage and the use of widely separated fields for different set of simulation parameters are therefore the optimal strategy to minimise the effect of cosmic variance at fixed total sky area.  Of the `Wide' fields, we use only the `W3' one given its large overlap with PAUS data and the presence of photo-z estimates from PAUS.

\subsection{Physics of the Accelerating Universe Survey}
\label{subsection:data_paus}

PAUS \cite{Benitez2009,marti14,Cabayol2019,Tonello2019,Cabayol2020} is a narrow-band photometric wide-field galaxy survey conducted at the William Herschel Telescope in the Observatorio del Roque de los Muchachos in La Palma (Canary Islands, Spain). The survey is carried out using the PAU camera (PAUCam \cite{Castilla2012,Castander2012,padilla16,Padilla2019}), a 18 CCDs camera that covers about 1 square degree in the sky. The 8 central CCDs cover about 1300 square arcminutes, excluding the gaps between CCDs, and they are those currently used for science. The average pixel scale of the camera is $0.265$ arcsec/pixel. The novelty of PAUCam resides in its set of 40 narrow-band filters spanning the range $4500-8500$ \AA\ with a width of 130 \AA\ (see figure \ref{fig:tortorelli_fig1}).\footnote{\href{https://www.pausurvey.org/paucam/filters/}{https://www.pausurvey.org/paucam/filters/}} Each filter slightly overlaps with the adjacent ones. PAUCam is also equipped with a standard set of optical broad-band filters (u, g, r, i, z, Y). The broad-band filters are placed on 6 trays that cover all the CCDs, while the narrow-band filters are arranged on 5 filter trays, with 8 filters per tray, located on the 8 central CCDs.\footnote{\href{https://www.pausurvey.org/paucam/detectors/}{https://www.pausurvey.org/paucam/detectors/}} Given the peculiar filter arrangement, in order to image a galaxy in all 40 narrow-bands, a specific observation pattern needs to be followed. After careful consideration, the PAUS collaboration decided to cover the PAUS fields with a spiral pattern. The target fields covered by PAUS are the COSMOS field \cite{Scoville2007}, the CFHTLS `W1', `W3' and `W4' fields and the KiDS/GAMA G09 field. In our work, we only use data overlapping with the CFHTLS `W3' field, with a total area on the sky of $\sim 30$ square degrees. We use already reduced PAUS images from the production id 813. These images have been processed by the PAUS collaboration using the \textsc{NIGHTLY} pipeline \cite{Tonello2019}.

The main aim of PAUS is to take advantage of the resolution of narrow-band observations to map the large scale structure (hereafter, LSS) of the universe in three dimensions up to $\mathrm{i_{AB}} < 22.5$ with a quasi-spectroscopic redshift precision of $\sigma(\mathrm{z})/(1+\mathrm{z}) \sim 0.0035$. The LSS mapping is carried out using galaxy clustering, intrinsic alignment of galaxies and Lyman $\alpha$ intensity mapping, of which forecast and results from the PAUS collaboration are already available \cite{stothert18,Johnston2021,Renard2021}. These measurements require a precision in the photometric redshift estimation that need new approaches and codes to be developed. Work in this direction has been carried out with a modified version of the BCNZ2 code \cite{Eriksen2019,Alarcon2021} and with the use of machine learning methods \cite{Eriksen2020,Soo2021}. These works were able to provide the photometric redshift precision that PAUS aimed for.

\section{Galaxy Population Model}
\label{section:gal_pop_model}

In this section, we provide a brief description of the galaxy population model we adopt in this work and of the image simulator UFig. A more thorough description can be found in \cite{herbel17,fagioli18,Tortorelli2020} (see in particular figure 1 of \cite{Tortorelli2020}). UFig \cite{berge13,bruderer16} is a code developed to simulate wide-field optical astronomical images. It generates catalogues of galaxies and then renders their single S\'ersic light profiles on a pixelated grid adding observational and instrumental effects. Its run-time is less than one minute for a $0.25$ square degrees optical image, roughly of the same order of magnitude as commonly used analysis software such as \textsc{Source Extractor} (hereafter, SE) \cite{bertin96}. This is achieved through optimised code and the use of simple yet realistic parametric models. The code has been used both for cosmological and galaxy evolution applications \cite{Tortorelli2018b,Tortorelli2020,bruderer16,herbel17,Kacprzak2020,Fagioli2020}.

In the galaxy population model, we sample blue and red galaxies from two different redshift-dependent B-band LFs. Their functional form is given by the Schechter function \cite{schechter76}
\begin{equation}
\Phi (\mathrm{M,z}) \mathrm{dM} = 0.4 \ln{(10)} \phi^*(z) 10^{0.4 (\mathrm{M^*(z) - M}) (\alpha(z) + 1)} \exp{{\left[ -10^{0.4 (\mathrm{M(z)^* - M})} \right]}} \mathrm{dM},
\end{equation}
where M is the absolute magnitude. The parameters describing the normalisation and the bright-end cut-off  of the LF, $\mathrm{M^*}$ and $\phi^*$, are kept as free parameters, while the faint-end slopes $\alpha$ are fixed to their observed low-redshift values \cite{beare15}. $\mathrm{M^*}$ and $\phi^*$ are, in turn, parametrised as linearly dependent on redshift
\begin{equation}
\begin{split}
\mathrm{M^*(z)} &= \mathrm{M^*_{slope}\ z + M^*_{intcpt} }, \\
\mathrm{\ln{\phi^*(z)}} &= \mathrm{\ln{\phi^*_{amp}} + \phi^*_{exp}\ z}, \\
\end{split}
\end{equation}
where $\mathrm{M^*_{slope}}$ and $\mathrm{M^*_{intcpt}}$ are the slope and the intercept of the linear evolution of $\mathrm{M^*}$ with redshift, while $\mathrm{\phi^*_{amp}}$ and $\mathrm{\phi^*_{exp}}$ are the amplitude and the exponential decay rate of $\phi^*$ with redshift that we linearise in log-space. From the LFs, UFig computes the comoving number of objects and assigns redshifts and absolute magnitudes in the B-band to galaxies. 

In this work, galaxies are randomly distributed on the sky within the specific simulated image. The absence of an accurate angular correlation function might affect the completeness of faint sources, especially in highly clustered areas \cite{zhang2015}. An extension of UFig containing a clustering prescription is in preparation \cite{Berner2021}. In \cite{Berner2021}, the authors make use of a sub-halo abundance matching approach to spatially distribute galaxies in 3D space in a realistic manner. The clustering of galaxies will allow to better control systematics and can be used as a separate cosmological probe.

Galaxy physical sizes are drawn from a log-normal distribution \cite{Shen2003} with mean $\mu_{\mathrm{phys}}$ and standard deviation $\sigma_{\mathrm{phys}}$. The mean $\mu_{\mathrm{phys}}$ of this distribution is linearly dependent on the absolute magnitude $\mathrm{M_B}$ drawn from the LF
\begin{equation}
\mu_{\mathrm{phys}} \left( \mathrm{M} \right) = \mathrm{r_{50,slope}^{phys}}\ \mathrm{M} + \mathrm{r_{50,intcpt}^{phys}}.
\end{equation}
Angular sizes are then derived from the physical sizes according to the galaxy redshift and the chosen cosmology. Galaxy properties are collected in a catalogue and then rendered on a pixelated grid according to a S\'ersic profile. The S\'ersic indices distributions follow the prescription in \cite{berge13}. The light profile is then convolved with the point spread function (hereafter, PSF) of the image that we model as a circular Moffat profile, whose parameters are estimated from observed data. Noise is then added to the image as a Gaussian distribution with mean and standard deviation estimated from the observed data. Correlated noise is introduced using a Lanczos resampling \cite{duchon79} of order 3. Stars are also simulated on the image using the Besan\c{c}on model \cite{robin03} of Milky-Way stars distribution.

The SEDs of galaxies $\mathrm{f(\lambda)}$ in the galaxy population model are constructed as linear combinations of templates $\mathrm{T_i(\lambda)}$ weighted by coefficients $\mathrm{c_i}$:
\begin{equation}
\mathrm{f(\lambda)} = \mathrm{\sum_{i=1}^{5} c_i T_i}.
\end{equation}
This is empirically motivated by SDSS data \cite{herbel17}. The $\mathrm{c_i}$ coefficients are jointly drawn from a Dirichlet distribution of order $5$, which is parametrised by five parameters $\mathrm{a_i}$. These are the spectral coefficients that we aim to constrain in this work. The coefficients $\mathrm{c_i}$ are used to construct the spectrum that is later rescaled to match its corresponding absolute magnitude $\mathrm{M_B}$. Since they are used to rescale the spectrum, only the $\mathrm{c_i}$ ratios are relevant, so the sum is arbitrarily chosen to be 1 to accommodate for the property of a Dirichlet distribution. The spectral coefficients $\mathrm{a_{i}}$ evolve with redshift $\mathrm{z}$ according to 
\begin{equation}
\mathrm{ a_i \left( z \right) = \left( a_{i,0} \right)^{1 - z / z_1} \times  \left( a_{i,1} \right)^{z/z_1}}.
\end{equation}
$\mathrm{a_{i,0}}$ samples the galaxy population at redshift $\mathrm{z} = 0$, while  $\mathrm{a_{i,1}}$ at redshift $\mathrm{z=z_1 >0}$. We use $\mathrm{z_1 = 1}$ in this work. The spectral coefficients are different for red and blue galaxies. The templates $\mathrm{T_i(\lambda)}$ are the five \textsc{kcorrect} templates presented in the rightmost column of figure 4 in \cite{blanton07}. They are constructed using Single Stellar Populations (SSP) from the Bruzual and Charlot stellar evolution synthesis models \cite{bruzual03}. They use a Chabrier Initial Mass function (IMF, \cite{Chabrier2003}), a solar value for the $\alpha$-to-iron ($\mathrm{[\alpha/Fe]}$) ratios and a grid of metallicities and ages in the range $0.0001 \le \mathrm{Z} \le 0.05$ and $\mathrm{1\ Myr \le t_{age} \le 13.75\ Gyrs}$, respectively. In addition, 35 templates from MAPPINGS-III \cite{Allen2008} are used to model the ionised gas emission. The 5 \textsc{kcorrect} templates are built with data up to redshift $z \sim 1.5$. They are derived via non-negative matrix factorisation trained on photometric and spectroscopic data from SDSS \cite{Eisenstein2001,Strauss2002}, 2MASS \cite{Jarrett2000}, GALEX \cite{Martin2005}, DEEP2 \cite{Davis2003} and GOODS \cite{Giavalisco2004}. The first template (see the rightmost column of figure 4 in \cite{blanton07}) represents a class of galaxies that had the last episode of star-formation around a lookback time of $0.5-1$ Gyrs and since then is passively evolving. The second template represents a class of galaxies that are strongly forming stars, as we can see from the rise in the ultraviolet emission, as well as from the emission lines. The third and fourth templates represent typical star-forming and passive galaxies, respectively. The last template is that of the class of post-starburst galaxies, as suggested by the strong Hydrogen absorption and spectrum that resembles that of an A type star. The rest-frame SEDs are then redshifted to the galaxy redshift $\mathrm{z}$ and position and wavelength-dependent extinction from the Milky-Way is applied. The observed spectrum is then integrated in the specific waveband probed to obtain the apparent magnitude of the galaxy in the AB photometric system.

\section{Method}
\label{section:method}

This section describes the method we use to constrain the galaxy population model parameters. We start by describing how we perform the photometric measurements on broad-band and narrow-band observed and simulated data. In order for the forward-modelling approach to be successful, we perform the same analysis steps and selection cuts on observed data and simulated data. The measurements can be divided in three main blocks (see the flowchart in figure 2 of \cite{Tortorelli2018b}): forced photometry on broad-band CFHTLS data to generate the detection images, forced photometry on PAUS narrow-band data and PAUS plus CFHTLS final catalogue combination. The ABC inference scheme we use is described in \cite{Tortorelli2020}. The main differences we implement are the initial prior and the use of new distances that take advantage of the rich wavelength information PAUS provides.

\subsection{CFHTLS image simulations and analysis}
\label{subsection:cfhtls_sims_analysis}

The detection of sources on narrow-band images is complicated by the narrowness of the wavelength window, which results in shallower data compared to broad-band ones at similar exposure times. This condition is further amplified if, e.g., a galaxy absorption line falls inside a narrow-band filter. It is therefore customary in narrow-band photometry to rely upon an external detection catalogue from an overlapping broad-band survey. In this work, we use CFHTLS images for this task. Four main reasons drive the use of CFHTLS images as opposed to PAUS images for the detection. The first is that CFHTLS is the dataset on which we constrained the galaxy population model parameters in \cite{Tortorelli2020} that we use as priors for this work. The second is that the PAUS data we use are from the `W3' field which is a CFHTLS `Wide' field. The third is that despite having higher transmission than broad-band filters, PAUS images are single exposures with an exposure time of a factor $\gtrsim 30$ shallower than the total exposure time of CFHTLS co-adds. This results in nearly two magnitudes difference in completeness \cite{cuillandre12,Benitez2009,marti14}. The last reason is the image survey strategy. Since every CCD of the PAU camera has a different set of 5 narrow-band filters, in order to image an object in all 40 narrow-bands, the camera follows a spiral dithering pattern. This makes the creation of a combined detection image extremely hard because the 40 narrow-bands do not overlap perfectly on the sky, as opposed to what CFHTLS co-added images do in their 5 bands.

For both simulating and analysing CFHTLS images, we need to obtain the survey instrumental parameters. These are gain, saturation, pixel scale, exposure time, magnitude zero-point, background noise and PSF FWHM for each image in the 5 wavebands. Gain, saturation, pixel scale, exposure time and magnitude zero-point are read from the CFHTLS images headers. We estimate the PSF FWHM of each image by using `GAIA DR2' stars \cite{gaia,gaiadr2} with g-band magnitude in the range $\mathrm{19 < m < 20}$ to have high signal-to-noise non-saturated stars. We cut stamps of $30 \times 30$ pixels around the stars and we fit a 2D circular Moffat profile to obtain its best-fit estimate of the $\beta$ parameter \cite{Moffat1969}. $\beta$ is linked to the FWHM via $\mathrm{FWHM = 2 \gamma \sqrt{2^{1/\beta}-1} }$. The mean values from all stars in the image are taken as final estimates of the $\mathrm{FWHM}$ and of the $\beta$. We measure the background noise mean and standard deviation from CFHTLS sigma-clipped observed data after masking galaxies through the SE segmentation image.

We perform the photometric measurements on observed data and simulations with SE. The SE configuration file we use in this work is the one described in appendix A of \cite{Tortorelli2020}. To avoid bias of the photometric measurements by star halos and reflections \cite{Erben2009,Gwyn2012,Heymans2012}, we mask stars as in \cite{Tortorelli2020}. These masks are provided to SE as weight images, allowing the software to ignore sources falling inside star halos or bleed trails. We run SE in dual-image mode to perform forced photometry. The dual-image mode requires the creation of a detection image from which SE estimates isophotal apertures. The detection image is created following the prescription in \cite{coe06} and references therein. It involves stacking the $5$-band CFHTLS images after normalising them by the rms of the background noise. We then apply the following selection cuts on the SE output catalogues of galaxy properties for every waveband:
\begin{equation}
\begin{split}
\mathrm{FLAGS} &< 4, \\ 
\mathrm{CLASS\_STAR} &< 0.9, \\
\mathrm{MAG\_ISO} &< 99, \\
\mathrm{MAG\_AUTO} &< 99, \\
\mathrm{FLUX\_RADIUS} &< 30.
\end{split}
\end{equation}
These cuts allow us to avoid spurious detections and sample contamination, to separate galaxies from stars and to keep galaxies with reliable photometry. We also apply PSF and aperture corrections to magnitudes following \cite{coe06}. The aperture and PSF-corrected magnitudes are used to compare observations and simulations via distance metrics. For simulated images, we also match the SE output catalogue with the UFig input one to assign redshifts and intrinsic properties to each simulated objects. 

Differently from \cite{Tortorelli2020}, we do not split the CFHTLS `Wide' images in patches of $4000 \times 4000$ pixels, but we use the original size images. This is motivated by the need to have large contiguous areas to maximise the number of sources with 40 narrow-bands measurements, given the spiral shape strategy of PAUS. We simulate the CFHTLS images in the 5 broad-bands. The noise and the PSF of the simulated images are position-independent. We check that the variation in background and PSF throughout the field was not impacting the photometry of the objects to more than few percent. We vary the noise seed for each simulated image, such that simulations of the same patch of the sky with different galaxy population model parameters have different galaxy positions and properties. Given the simplistic treatment of stars in simulations, we do not create masks for the CFHTLS simulated images, but we use the ones for survey data. We assume that simulated saturated stars and artefacts lie in the same position on the sky as real ones and we therefore simulate only stars with magnitude $\mathrm{m > 15}$.

\subsection{PAUS image simulations and analysis}
\label{subsection:paus_sims_analysis}

\begin{figure}[t!]
\centering
\includegraphics[width=14cm]{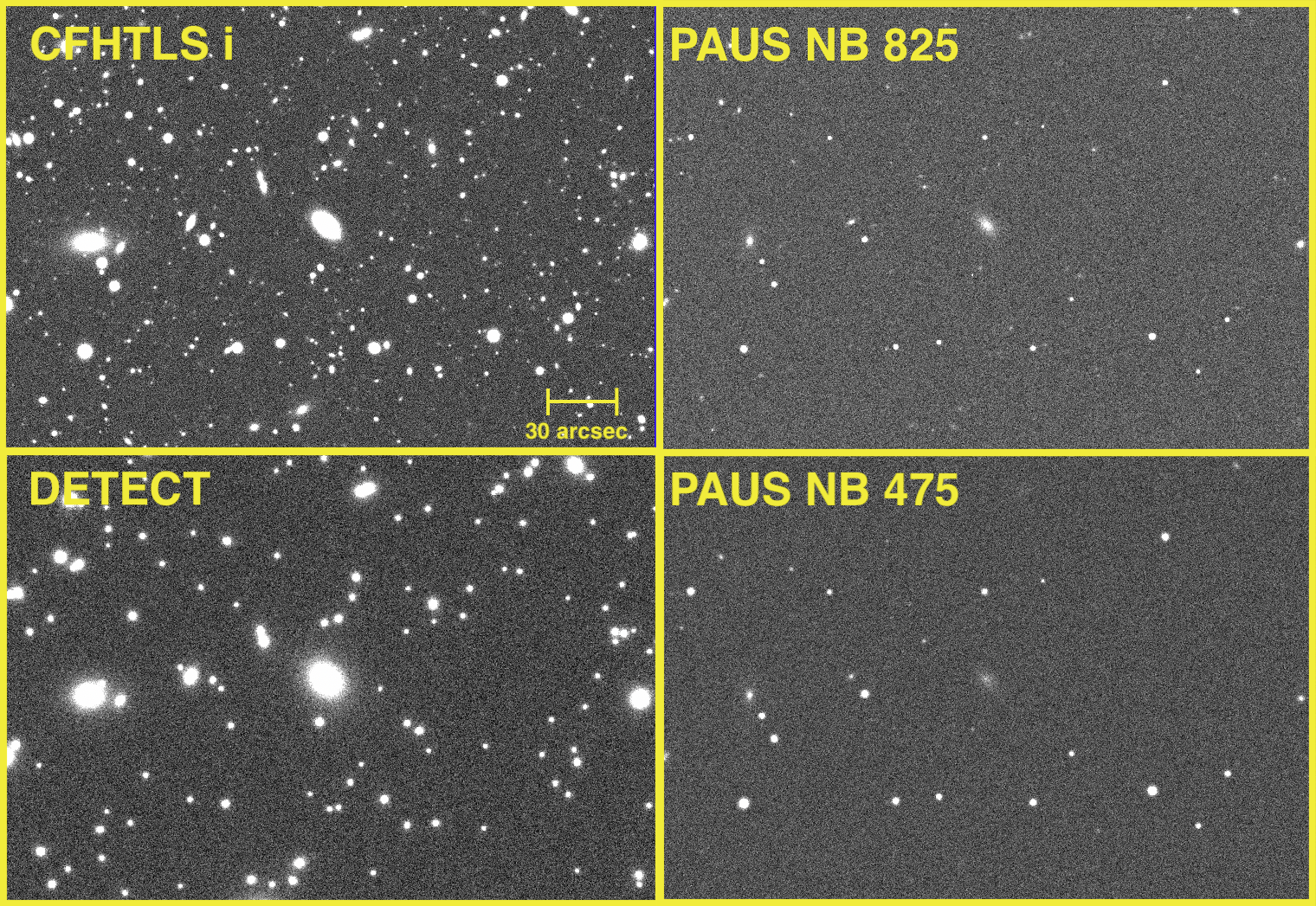}
\caption{Example cut-outs ($\mathrm{3\ arcmin \times 3\ arcmin}$) of CFHTLS and PAUS simulated images. The upper left panel shows a simulated CFHTLS image in the i-band. The lower left panel shows a cut-out of the corresponding detection image for PAUS forced photometry. The right panels show the overlapping simulated PAUS images in the narrow-band centred at 8250 \AA\ and 4750 \AA, respectively. The detection images contain only sources having $\mathrm{i'} \le 22.5$, as explained in section \ref{subsection:paus_sims_analysis}, while the simulated CFHTLS and PAUS images have no magnitude cut.}
\label{fig:tortorelli_fig2}
\end{figure}

Forced photometry is the preferred approach for narrow-band imaging given the difficulty in detecting sources. Usually, a deeper catalogue of broad-band detections is provided. However, SE does not accept a detection catalogue, but we need to provide detection and measurement images. Additionally, the detection and the measurement images in SE must have the same pixel size and pixel scale. Therefore, we need to create a detection image using UFig. Since PAUS is complete up to $\mathrm{i'} \le 22.5$, we select as broad-band detections only those galaxies from CFHTLS observations and simulations that satisfy this magnitude cut. These galaxies are the ones for which the narrow-band photometry is performed and therefore they are also the ones used to create the detection images. CFHTLS and PAUS simulated images, instead, contain all sources without any magnitude cut, as they are a realistic representation of observed data.

For each given CFHTLS image, we first find all the overlapping PAUS fields (roughly $\sim 30$) that we intend to analyse. For each PAUS field, the instrumental properties are measured as in section \ref{subsection:cfhtls_sims_analysis}, although in this case the PSF FWHMs are taken from the PAUS database.\footnote{\href{http://paudm.pau.pic.es}{http://paudm.pau.pic.es}} The magnitude zero-point for each narrow-band is defined as $\mathrm{zpt = 26 - 2.5 \log{(zpt\_nightly)}}$. $\mathrm{zpt\_nightly}$ is the nightly zero-point estimated with the PAUS \textsc{NIGHTLY} pipeline \cite{Tonello2019}. This is a multiplicative factor that is used to obtain calibrated fluxes and is different for every band and observing night. We simulate PAUS images by using as input galaxy properties those that are used to simulate CFHTLS images. The effect of scattered light present in real data is not simulated by UFig. 

We then create the detection images with UFig for both PAUS observed data and simulations. The detection images have the same pixel sizes and pixel scales as the PAUS images. The noise properties are instead from the overlapping CFHTLS image in the i-band. Similarly, the galaxy properties are taken from the SE output catalogues of CFHTLS in the i-band. Sizes are rescaled according to the ratio of the pixel scales between CFHTLS and PAUS. 

Since UFig renders galaxies according to S\'ersic light profiles, to be consistent between data and simulations, we need to estimate the S\'ersic indices from CFHTLS images. We do that by providing SE during the dual-image mode with a PSF model for each CFHTLS image. The same stars used to measure the PSF FWHM are median combined to obtain the model PSF we provide SE with. 

The same detection image constructed from CFHTLS broad-band properties is then used for all the 40 narrow-bands. This ensures that the apertures for each galaxy are consistent throughout the whole narrow-band spectrum, fixed by the broad-band image. Figure \ref{fig:tortorelli_fig2} shows an example of simulated CFHTLS $\mathrm{i'}$-band image, detection image and overlapping PAUS images in two different narrow-bands, covering the blue and red part of the optical spectrum.

We run SE in dual-image mode. Aperture and PSF corrections are applied to the output SE catalogues. We apply the same quality cuts as in \ref{subsection:cfhtls_sims_analysis}. All the measurements from the 40 narrow-band SE catalogues are initially split in several single-band catalogues. All the single-band catalogues are combined into a single master catalogue. We assign to each galaxy an ID from the CFHTLS SE output catalogue, as well as the information about the measured narrow-band. This way we can match the objects and create the final catalogue containing the 40 narrow-band fluxes for each galaxy by simply indexing objects by keywords instead of sky coordinates. 

Since the different narrow-bands have different zero-points, the calibrated fluxes are obtained using the `MAG\_AUTO' SE keyword as $\mathrm{f_{cal} = 10^{(MAG\_AUTO - 26) / (-2.5)}}$. For objects having multiple exposures in a single narrow-band, the final calibrated fluxes and magnitudes are obtained through a weighted mean. The weights are given by the signal-to-noise ratio of each measurement, $\mathrm{S/N = `FLUX\_AUTO' / `FLUXERR\_AUTO'}$, such that the weights are approximately proportional to the exposure time of the observation. `FLUX\_AUTO' and `FLUXERR\_AUTO' are the uncalibrated fluxes and corresponding errors measured with SE.

\begin{table}
\centering
\resizebox{\textwidth}{!}{
\begin{tabular}{p{4cm} p{4cm} p{3cm} p{4cm} p{3cm}}
\hline
\hline
\textbf{Parameter} & \textbf{Tortorelli+20 Distribution} & \textbf{Tortorelli+20 Prior} & \textbf{This work Distribution} & \textbf{This work Prior}\\
\hline
$\alpha$ (blue) & Fixed value & -1.3 & Fixed value & -1.3 \\
\hline
$\alpha$ (red) & Fixed value & -0.5 & Fixed value & -0.5 \\
\hline
$\mathrm{M^*_{B,slope}}$ (blue) & Multivariate Normal & $\mu = -9.44 \times 10^{-1}$,  $\sigma^2 = 8.29\times10^{-1}$ & Multivariate Normal & $\mu = -6.44 \times 10^{-1}$,  $\sigma^2 = 5.08 \times10^{-1}$\\
\hline
$\mathrm{M^*_{B,slope}}$ (red) & Multivariate Normal & $\mu = -7.33 \times 10^{-1}$, $\sigma^2 = 5.30\times10^{-1}$ & Multivariate Normal & $\mu = -7.14 \times 10^{-1}$,  $\sigma^2 = 5.09 \times10^{-1}$\\
\hline
$\mathrm{M^*_{B,intcpt}} - 5 \log{h_{70}} $ (blue) & Multivariate Normal & $\mu = -2.041 \times 10^{1}$,  $\sigma^2 = 3.312\times10^{-1}$ & Multivariate Normal & $\mu = -2.05 \times 10^{1}$,  $\sigma^2 = 2.35 \times10^{-1}$\\
\hline
$\mathrm{M^*_{B,intcpt}} - 5 \log{h_{70}} $ (red) & Multivariate Normal & $\mu = -2.035 \times 10^{-1}$, $\sigma^2 = 2.968\times10^{-1}$ & Multivariate Normal & $\mu = -2.03 \times 10^{1}$,  $\sigma^2 = 2.75 \times10^{-1}$\\
\hline
$\mathrm{\phi^*_{exp}}$ (blue) & Multivariate Normal & $\mu = -5.66 \times 10^{-2}$,  $\sigma^2 = 9.96\times10^{-2}$ & Multivariate Normal & $\mu = -7.69 \times 10^{-2}$,  $\sigma^2 = 9.65 \times10^{-2}$\\
\hline
$\mathrm{\phi^*_{exp}}$ (red) & Multivariate Normal & $\mu = -6.97 \times 10^{-1}$, $\sigma^2 = 9.21\times10^{-1}$ & Multivariate Normal & $\mu = -7.00 \times 10^{-1}$,  $\sigma^2 = 6.74 \times10^{-1}$\\
\hline
$\mathrm{\ln{\phi^*_{amp}}}$ / 10$^{-3}$ h$^3_{70}$ Mpc$^{-3}$ mag$^{-1}$ (blue) & Multivariate Normal & $\mu = -5.28 \times 10^{0}$, $\sigma^2 = 4.1 \times 10^{-1}$ & Multivariate Normal & $\mu = -5.54 \times 10^{0}$,  $\sigma^2 = 1.80 \times10^{-1}$\\
\hline
$\mathrm{\ln{\phi^*_{amp}}}$ / 10$^{-3}$ h$^3_{70}$ Mpc$^{-3}$ mag$^{-1}$ (red) & Multivariate Normal & $\mu = -5.28 \times 10^{0}$, $\sigma^2 = 6.5 \times 10^{-1}$ & Multivariate Normal & $\mu = -5.38 \times 10^{0}$,  $\sigma^2 = 5.47 \times10^{-1}$\\
\hline
$\mathrm{r_{50,slope}^{phys}}$ & Multivariate Normal & $\mu = -2.4 \times 10^{-1}$, $\sigma^2 = 9.8\times10^{-6}$ & Multivariate Normal & $\mu = -2.42 \times 10^{-1}$,  $\sigma^2 = 9.88 \times10^{-6}$\\
\hline
$\mathrm{r_{50,intcpt}^{phys}}$ & Uniform & [-2, 4] & Multivariate Normal & $\mu = 8.48 \times 10^{-1}$,  $\sigma^2 = 5.89 \times10^{-2}$\\
\hline
$\mathrm{\sigma_{phys}}$ & Multivariate Normal & $\mu = 5.7 \times 10^{-1}$, $\sigma^2 = 1.9 \times 10^{-5}$ & Multivariate Normal & $\mu =5.67 \times 10^{-1}$,  $\sigma^2 = -6.44 \times10^{-5}$\\
\hline
$\mathrm{a}_{i,0}$ (blue) & Dirichlet $\times$ Uniform & [1., 1., 1., 1., 1.] $\times$ [5, 15] & Dirichlet $\times$ Uniform & [1., 1., 1., 1., 1.] $\times$ [5, 15]\\
\hline
$\mathrm{a}_{i,0}$ (red) & Dirichlet $\times$ Uniform & [1., 1., 1., 1., 1.] $\times$ [5, 15] & Dirichlet $\times$ Uniform & [1., 1., 1., 1., 1.] $\times$ [5, 15]\\
\hline
$\mathrm{a}_{i,1}$ (blue) & Dirichlet $\times$ Uniform & [1., 1., 1., 1., 1.] $\times$ [5, 15] & Dirichlet $\times$ Uniform & [1., 1., 1., 1., 1.] $\times$ [5, 15]\\
\hline
$\mathrm{a}_{i,1}$ (red) & Dirichlet $\times$ Uniform & [1., 1., 1., 1., 1.] $\times$ [5, 15] & Dirichlet $\times$ Uniform & [1., 1., 1., 1., 1.] $\times$ [5, 15]\\
\hline
\end{tabular}
}
\caption{Prior range of the parameters used to simulate CFHTLS and PAUS images for red and blue galaxies. In \cite{Tortorelli2020}, the LF and size parameters are drawn from two Multivariate Normal distributions, except for $\mathrm{r_{50,intcpt}^{phys}}$ which has a uniform distribution, while the spectral coefficients are drawn from Dirichlet distributions of order five. In this work, we obtain the prior for the LF and size parameters by resampling the posterior from \cite{Tortorelli2020} using a Multivariate Gaussian. The covariance of this Gaussian is estimated from the posterior distribution in \cite{Tortorelli2020}. The prior for the spectral coefficients is the same as \cite{Tortorelli2020}.}
\label{table:tortorelli_priortable}
\end{table}

\subsection{ABC inference scheme}
\label{subsection:abc_scheme}

Likelihood-free inference methods need to be used in cases where the likelihood function is either unknown or there is no clear empirical likelihood that can be calculated for simulation-based models. ABC \cite{akeret15} has been proven to be a robust likelihood-free method for both cosmology and galaxy evolution studies. The idea is to approximate the true posterior by iteratively restricting the prior space on the base of a distance metric between observed $x$ and simulated data $y$:
\begin{equation}
p \left ( \theta | y \right ) \simeq p \left ( \theta | \rho \left(x,y \right ) \le \epsilon \right),
\end{equation}
where $\theta$ is the set of model parameters, $\rho$ is the distance metric and $\epsilon$ is a specified threshold.

The prior we adopt is described in table \ref{table:tortorelli_priortable}, while the distance metrics we define are described in section \ref{subsection:distance_metrics}. In \cite{Tortorelli2020}, we built the prior for the LF parameters using measurements from \cite{beare15} enlarging the uncertainties by a factor of 10. Sizes are obtained from the Great-3 dataset \cite{mandelbaum14} enlarging the uncertainties by a factor of 3. We choose these factors in \cite{Tortorelli2020} to create a large prior space that was as conservative as possible, albeit centred around observed sizes and luminosity function distributions. Larger values were not considered since they lead to unphysical parameter values. The spectral coefficients are instead drawn from Dirichlet distributions of order five. The prior for the LF and size parameters in this work is given by the posterior from \cite{Tortorelli2020} resampled using a Multivariate Gaussian with covariance estimated from the posterior distribution itself. The prior for the spectral coefficients is the same as in \cite{Tortorelli2020}. This is motivated by the fact that CFHTLS data saturates the available information to estimate LF parameters, while they leave the spectral coefficients only broadly constrained. The addition of PAUS data is expected to break this degeneracy by providing a dataset that is sensitive mostly to the spectral coefficients, rather than the LF parameters. Therefore, it is safe to assume that the new LF and size posterior will lie in the parameter space that has already been constrained by CFHTLS data.

In this work, we use the ABC inference scheme described in figure 4 of \cite{Tortorelli2020}. It consists of multiple classical Rejection ABC steps \cite{Tavare1997}, where the posterior from an iteration constitutes the prior for the following one. In \cite{Tortorelli2020}, we increased the size of the dataset per set of parameters at each iteration. In this implementation, we keep the dataset size fixed for operational reasons. This is motivated by the large computational time required to simulate narrow-band images. Indeed, for a single ABC iteration where we sample $10^4$ points from the parameter space, we need to simulate $\sim 6 \times 10^{5}$ narrow band images. Following \cite{Tortorelli2020}, at each iteration T, we evaluate the galaxy population model on $\mathrm{N} = 10^4$ parameters $\theta^{*}_{\mathrm{i,T}}$ drawn from the prior. The approximate Bayesian posterior at iteration T is given by the samples having distance metrics less than a defined threshold $\epsilon$. The threshold is given by the $\mathrm{q}=10$-th percentile value for each distance metric. 

During the first iteration, we evaluate all the distance metrics in section \ref{subsection:distance_metrics}, while for the following ones we only evaluate the one that gives the most stringent constraints on all parameters in the first iteration. This distance metric is used both to estimate the posterior at T and to resample the posterior as prior for the consecutive iteration $\mathrm{T+1}$. At each iteration, we keep only the $\mathrm{N_{q,T}}$ samples having distance metric below $\epsilon_{\mathrm{T}}$, $\{ (\theta^{(\mathrm{T})}, \rho^{(\mathrm{T})}) \} = \{ (\theta_{\mathrm{i,T}}, \rho_{\mathrm{i,T}}) | \rho_{\mathrm{i,T}} \le \epsilon_{\mathrm{T}}, \mathrm{i=1,...,N_{q,T}} \}$ as approximated posterior. Then, the $\mathrm{N'} = 10^4$ new parameter sets $\theta^{*}_{\mathrm{i,T+1}}$ that constitute the prior for iteration $\mathrm{T+1}$ are drawn by resampling the approximated posterior. The resampling is carried out following the procedure highlighted in \cite{Moser2021}. We create a model of the posterior using the \textsc{scikit-learn} \cite{scikit-learn} implementation of a Gaussian mixture model (GMM), where we use $\mathrm{n=20}$ Gaussians to fit the distribution. Before fitting, we Gaussianise the model parameter space to make it more suitable to be fitted by the GMM. We do that by applying a Gaussian inverse-CDF transform. We draw the $\mathrm{N'} = 10^4$ resampled points from this Gaussianised space and then invert back to the original space. We do not fix the number of total iterations, but we keep drawing from the updated posterior until the acceptance ratio of newly drawn samples is $\mathrm{p_{acc}} < \mathrm{p_{acc,min}} = 10\%$.

\subsection{Distance Metrics}
\label{subsection:distance_metrics}

\begin{table}
\centering
\resizebox{\textwidth}{!}{
\begin{tabular}{p{8cm} p{2cm} p{3cm}}
\hline
\hline
\textbf{Distance Metric} & \textbf{Used Data} & \textbf{Label}\\
\hline
Absolute difference in the number of detected galaxies & CFHTLS & d$_1$ \\
\hline
Fraction difference in the number of detected galaxies & CFHTLS & d$_2$ \\
\hline
MMD distance on $\mathrm{u^{*}}$, $\mathrm{g'}$, $\mathrm{r'}$, $\mathrm{i'}$, $\mathrm{z'}$ band properties & CFHTLS & d$_{\{3,...,6\}}$ \\
\hline
Magnitude histogram distance on $\mathrm{u^{*}}$, $\mathrm{g'}$, $\mathrm{r'}$, $\mathrm{i'}$, $\mathrm{z'}$ bands separately & CFHTLS & d$_{\{7,...,11 \}}$ \\
\hline
Size histogram distance on $\mathrm{u^{*}}$, $\mathrm{g'}$, $\mathrm{r'}$, $\mathrm{i'}$, $\mathrm{z'}$bands separately & CFHTLS & d$_{\{12,...,16 \}}$ \\
\hline
Maximum value among all previously defined rescaled distances & CFHTLS & d$_{17}$= $\mathrm{max(\underline{d}_{\{1,...,16\}})}$ \\
\hline
Maximum value between the rescaled MMD distance on 5 bands and the rescaled absolute difference & CFHTLS & $\mathrm{ d_{\{18,...,21\}}}$= max$\mathrm{(\underline{d}_1,\underline{d}_{\{3,...,6\}}) }$ \\
\hline
Maximum value among the rescaled MMD distance and the rescaled magnitude histogram distance on 5 bands & CFHTLS & $\mathrm{d_{\{22,...,25\}}}$= max$\mathrm{(\underline{d}_{\{3,...,6\}},\underline{d}_{\{7,...,11 \}} ) }$ \\
\hline
Sum between the rescaled MMD distance on 5 bands and the rescaled absolute difference & CFHTLS & $\mathrm{ d_{\{26,...,29\}}}$= $\mathrm{\sum_{i=\{1,3,...,6\}}\underline{d}_i}$ \\
\hline
Sum between the rescaled MMD distance on 5 bands and the rescaled fractional difference & CFHTLS & $\mathrm{ d_{\{30,...,33\}}}$= $\mathrm{\sum_{i=\{2,3,...,6\}}\underline{d}_i}$ \\
\hline
PCA distance & PAUS & d$_{34}$ \\
\hline
MMD distance on PC coefficients & PAUS & d$_{35}$ \\
\hline
Sum among the rescaled MMD distance on 5 bands, the rescaled absolute difference, the rescaled PCA distance and the rescaled MMD distance on PC coefficients & CFHTLS + PAUS & $\mathrm{ d_{\{36,...,39\}}}$= $\mathrm{\sum_{i=\{1,3,...,6,34,35\}}\underline{d}_i}$ \\
\hline
Sum among the rescaled MMD distance on 5 bands, the rescaled fractional difference, the rescaled PCA distance and the rescaled MMD distance on PC coefficients & CFHTLS + PAUS & $\mathrm{ d_{\{40,...,43\}}}$= $\mathrm{\sum_{i=\{2,3,...,6,34,35\}}\underline{d}_i}$ \\
\hline
\end{tabular}
}
\caption{Table of distance metrics used in this work. $\mathrm{\underline{d_j}}$ refers to the rescaled value for the j-th distance metric. `MMD', `PCA' and `PC' stands for Maximum Mean Discrepancy, Principal Component Analysis and Principal Component, respectively.}
\label{table:tortorelli_distances}
\end{table}

Distance metrics are used to quantify the level of agreement between two datasets. The accurate choice of the distance metrics is of fundamental importance to ensure a physically meaningful ABC inference. The distance metrics may be sensitive to a specific aspect of the galaxy population model, e.g., the absolute difference in the number of galaxies is sensitive to the amplitude of the LF. Therefore, combining distance metrics is essential to ensure that we simultaneously constrain all the model parameters. In this work, we use part of the distance metrics defined in \cite{Tortorelli2020} and we add new ones to take advantage of the rich wavelength information provided by PAUS. Furthermore, we also add new distances for CFHTLS data. We report the used distance metrics in table \ref{table:tortorelli_distances}. We split the distance metrics in those evaluated on CFHTLS (or PAUS) only and those evaluated on the combined dataset.

\textbf{Absolute and fractional difference distances.} d$_1$ and d$_2$ are distance metrics that measure the absolute and fractional difference between the number of detected galaxies on observed data ($\mathrm{N_{data}}$) and simulations ($\mathrm{N_{sims}}$), respectively:
\begin{equation}
\begin{split}
\mathrm{d_1} &= |\ \mathrm{N_{data} - N_{sims}}\ |, \\
\mathrm{d_2} &= \frac{|\ \mathrm{N_{data} - N_{sims}}\ |}{\mathrm{N_{data}}}.\\
\end{split}
\end{equation}
These distance metrics are sensitive to the amplitude of the LF. Since PAUS galaxies are selected as a subsample of galaxies in CFHTLS having $\mathrm{i'} \le 22.5$, we only use CFHTLS data to estimate these distance metrics.

\textbf{Maximum Mean Discrepancy distance.} The distance metrics d$_{\{ 3,...,6\}}$ and d$_{35}$ are computed using the Maximum Mean Discrepancy (MMD) distance \cite{gretton08}. The MMD distance between two multi-dimensional probability distributions is defined as:
\begin{equation}
\mathrm{  d_{\{ 3,...,6\}}, d_{35} = \frac{1}{N(N-1)} \sum_{i,j} k(x_i,x_j) + k(y_i,y_j) - k(x_i,y_j) - k(y_i,x_j)    },
\end{equation}
with $\mathrm{N}$ being the size of the samples $\mathrm{x}$ and $\mathrm{y}$. $\mathrm{k}$ is a Gaussian kernel function of a pre-defined width $\sigma$
\begin{equation}
\mathrm{k (x_i,y_j) = \exp{\left( - \frac{\left \| x_i - y_j \right \|^2}{2\sigma} \right)}},
\end{equation}
where $\sigma$ is a free parameter \cite{gretton08}. Observed and simulated galaxy property distributions have been rescaled to have zero mean and unit variance. d$_{\{ 3,...,6\}}$ are the ten-dimensional MMD distances measured on the 5 CFHTLS wavebands: d$_3$ is the MMD distance between observed and simulated PSF corrected magnitudes and sizes, d$_4$ among PSF corrected magnitudes, sizes and colours, d$_5$ among PSF corrected magnitudes, sizes and flux fractions (ratio between SE `FLUX\_AUTO' in a waveband and the sum in all wavebands), d$_6$ among PSF corrected magnitudes, sizes, colours and flux fractions. d$_{35}$ is the MMD distance between the observed and simulated coefficients resulting from the scalar product between the Principal Components (hereafter, PCs) of PAUS observed data with the observed and simulated normalised narrow-band spectra (see MMD on Principal Component coefficients distance). The MMD distance on CFHTLS data mainly constrains the exponential cut-off of the LF and the size distribution, while the MMD distance on PAUS data mainly constrains the spectral coefficients distribution. However, none of them is sensitive to the amplitude of the LF.

\textbf{Histogram distance.} d$_{\{ 7,...,16\}}$ are the histogram distances, measured as absolute differences between the counts in 20 bins of two equally binned histograms for observed data $\mathrm{h_{data,i}}$ and simulations $\mathrm{h_{sims,i}}$:
\begin{equation}
\mathrm{d_{\{ 7,...,16 \}} = \sum_i  |\ h_{data,i} - h_{sims,i}\  | }.
\end{equation}
The bin number is motivated in appendix B of \cite{Tortorelli2020}. These distance metrics are measured for magnitudes (d$_{\{ 7,...,11\}}$) and sizes (d$_{\{ 11,...,16\}}$) in the 5 CFHTLS wavebands. They are sensitive to the amplitude of the LF, its exponential cut-off and the size distribution, but not to the spectral coefficients distribution.

\textbf{Principal Component Analysis distance}. d$_{34}$ is the Principal Component Analysis (hereafter, PCA) distance. We use this distance because it allows us to take full advantage of the spectral information provided by PAUS data. This distance metric is based on the comparison between the PCA performed on narrow-band observed and simulated spectra. We perform the PCA using its \textsc{scikit-learn} \cite{scikit-learn} implementation. The PCA requires that the input data are normalised and mean-subtracted to ensure a meaningful result. Therefore, we pre-process the narrow-band spectra by normalising them as described in section \ref{section:pca_nb_spectra}. We then use the same number of narrow-band observed and simulated spectra by randomly drawing $\mathrm{N = min(N_{observed}, N_{sims})}$ spectra from observed data and simulations. We compute the mean observed and simulated spectra as described in section \ref{section:pca_nb_spectra}. We subtract the mean spectra from the normalised ones to obtain the final mean subtracted normalised spectra. We use the latter spectra to perform the PCA with $\mathrm{n=5}$ components and obtain the corresponding PCs. We choose to use the same number of components as those of the spectral templates from \textsc{kcorrect}. We then introduce the so-called `mixing matrix' $\mathrm{M_{ij}}$,
\begin{equation}
\mathrm{M_{ij} = \int \phi_i(\lambda) \psi_j(\lambda) d\lambda}.
\end{equation}
This matrix is the scalar product between the observed data PCs $\mathrm{\phi_i(\lambda)}$ and the simulated PCs $\mathrm{\psi_j(\lambda)}$. Since the PCs are two sets of orthonormal basis, if there were perfect agreement, the mixing matrix would be diagonal $\mathrm{M_{ij} = \delta_{ij}}$. However, the presence of noise in the measurements, the number of objects and the analysis performed in the observed-frame prevent that from happening. In particular, since the analysis is performed in the observed-frame, the spectral features of the narrow-band spectra occurs at different redshifts, thereby smoothing the resulting PCs. This prevents high-order PCs to be signal rather than noise dominated. For this reason, we estimate d$_{34}$ using only the scalar product between the first two PCs. The PCA distance $\mathrm{d_{34}}$ is then defined as follows:
\begin{equation}
\mathrm{d_{34}} = (1 - \left| \mathrm{M_{11}} \right|) + (1 - \left| \mathrm{M_{22}} \right|) + \left( 1 - \left| \frac{ \prod \mathrm{M_{ii}} }{\det \mathrm{M_{ij}}} \right| \right), \quad \mathrm{i,j=\{1,2\}}.
\end{equation}
This distance would be 0 if there were perfect agreement. The first two terms measure the distance of the first two PCs, while the third terms allows us to weight the distance by decreasing the contribution from mixing matrices where the first two PCs are highly correlated. To avoid any bias due to the random selection of the $\mathrm{N}$ narrow-band spectra, we perform 100 realisations of the PCA distance. The final distance is the one among the 100 values that is the nearest to the median value of the realisations.

\textbf{MMD on PC coefficients distance.}  d$_{35}$ is the MMD distance on the PC coefficients. Given the PCs, the narrow-band spectra of observed $\mathrm{f}(\lambda)$ and simulated $\mathrm{f}^{'}(\lambda)$ galaxies can be decomposed as:
\begin{equation}
\begin{split}
\mathrm{f}(\lambda) &= \sum_{\mathrm{j}} \mathrm{m}_{\mathrm{j}} \phi_{\mathrm{j}}(\lambda), \\
\mathrm{f}^{'}(\lambda) &= \sum_{\mathrm{j}} \mathrm{n}_{\mathrm{j}} \psi_{\mathrm{j}}(\lambda),
\end{split}
\label{equa}
\end{equation}
where $\mathrm{m}_{j}$ and $\mathrm{n}_{j}$ are PC coefficients, i.e. the scalar products of the PCs with their normalised narrow-band spectra. $\phi_{\mathrm{j}}(\lambda)$ and $\psi_{\mathrm{j}}(\lambda)$ are the observed data and simulated PCs, respectively. We compute the PC coefficients for all 5 PCs and we compare the distribution of observed data and simulations using the MMD distance. As mentioned earlier, the MMD distance is computed on PC coefficients that have been scaled to have zero mean and unit variance.

\textbf{Maximum among rescaled distances.} One way to combine the constraining power of different distance metrics is to compute the maximum value among all the rescaled distance metrics:
\begin{equation}
\mathrm{d_{\{ 17,...,25 \}} = max(\underline{d}_{\{ l,...,k \}}) },
\end{equation}
where $\underline{\mathrm{d}}$ is the rescaled distance metric. We rescale each distance metric by dividing it for its mean value computed from the first 1000 samples of the first ABC iteration. This ensures that all rescaled distance metrics have the same order of magnitude.

\begin{figure}[t!]
\centering
\includegraphics[width=16cm]{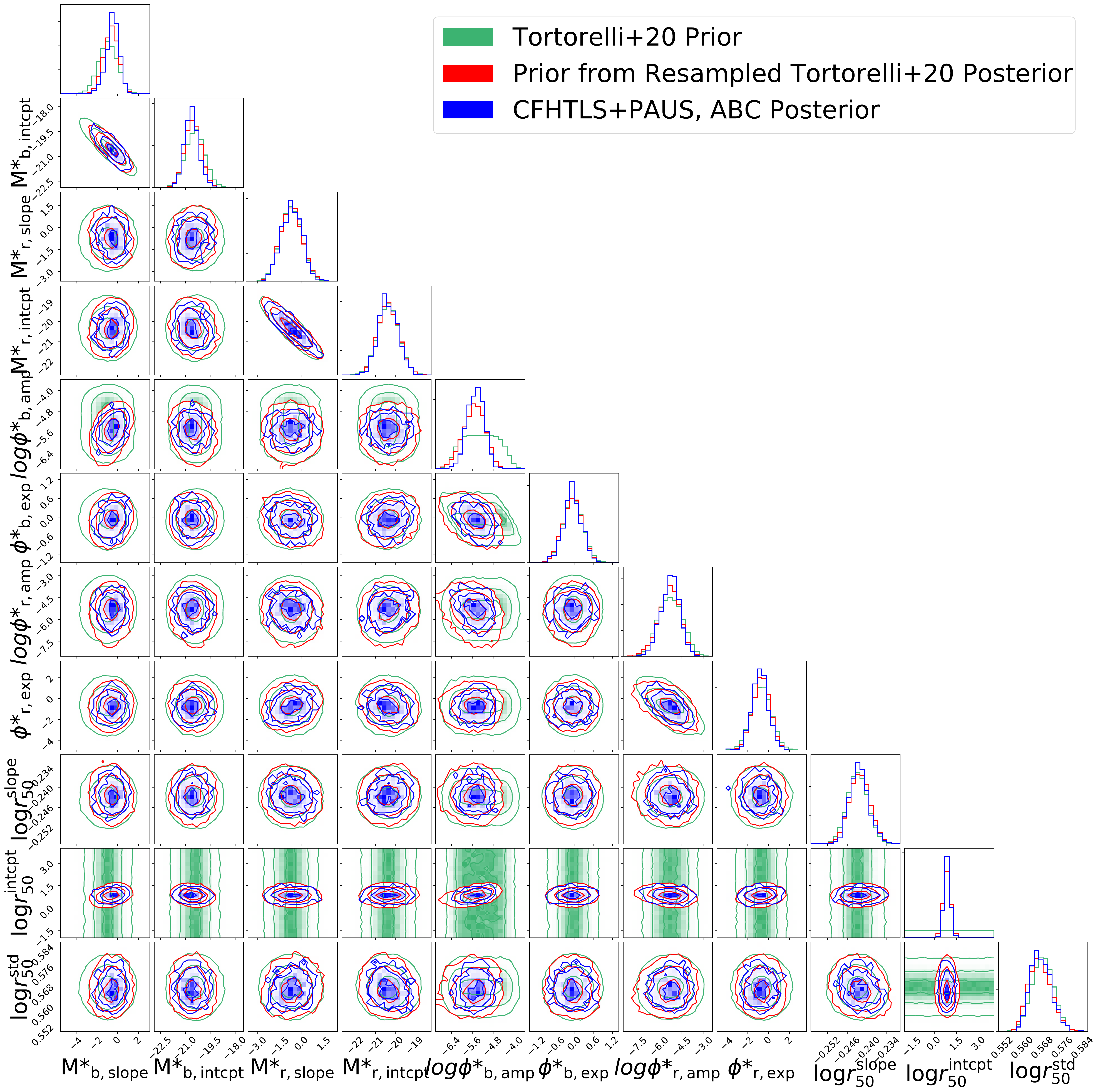}
\caption{Posterior distribution of the LF and size parameters for the ABC inference on CFHTLS and PAUS survey data. Green contours show the prior distribution from \cite{Tortorelli2020}. Red contours show the prior for this work obtained by resampling the posterior from \cite{Tortorelli2020} using a Multivariate Gaussian. Blue contours show the approximate Bayesian posterior distribution at the final iteration. The subscripts `b' and `r' refer to parameters for blue and red galaxies, respectively. The contour levels refer to the $\sim 39\%$, $\sim 86\%$ and $\sim 98\%$ confidence levels for a 2D Gaussian distribution.}
\label{fig:tortorelli_fig3}
\end{figure}

\textbf{Sum among rescaled distances.} Another way of combining the constraining power of multiple distance metrics is to sum their rescaled values:
\begin{equation}
\mathrm{d_{\{ 26,...,33 \}}, d_{\{ 36,...,43 \}} = \sum_{i=\{ l,...,k \}} \underline{d}_i },
\end{equation}
where $\underline{\mathrm{d_i}}$ is the rescaled i-th distance metric. Both the maximum and the sum among the rescaled distance metrics are sensitive to all parameters of the galaxy population model.

\section{ABC inference on CFHTLS and PAUS data}
\label{section:abc_cfhtls_paus}

This section presents the result from the ABC inference on CFHTLS and PAUS survey data. We show the approximate Bayesian posterior distribution of our 31 parameters galaxy population model (8 LF parameters, 3 size parameters, 20 spectral coefficients) compared with the \cite{Tortorelli2020} prior and the prior for this work.

\begin{figure}[t!]
\centering
\includegraphics[width=16cm]{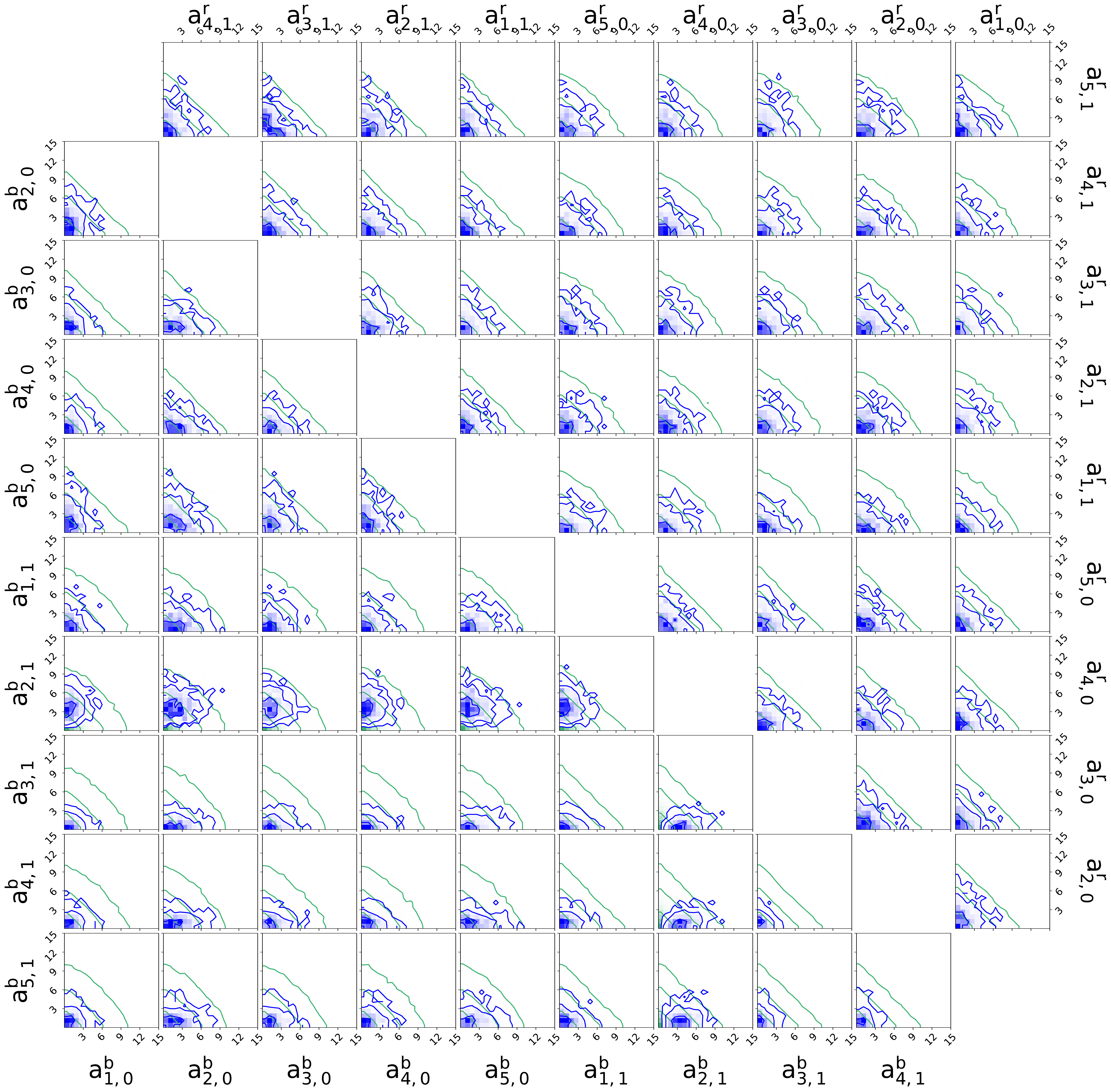}
\caption{Posterior distribution of the spectral coefficients for blue galaxies (lower corner plot) and red galaxies (upper corner plot) for the ABC inference on CFHTLS and PAUS survey data. Green contours show the prior distribution from \cite{Tortorelli2020}, which is also used as prior in this work. Blue contours show the approximate Bayesian posterior distribution at the final iteration. The subscript `b' and `r' refer to parameters for blue and red galaxies, respectively. The subscripts `$0$' and `$1$' for the five components refer to the coefficients at redshift $\mathrm{z = 0}$ and $\mathrm{z = 1}$. The contour levels refer to the $\sim 39\%$, $\sim 86\%$ and $\sim 98\%$ confidence levels for a 2D Gaussian distribution.}
\label{fig:tortorelli_fig4}
\end{figure}

As explained in section \ref{subsection:abc_scheme}, the ABC scheme we adopt is taken from \cite{Tortorelli2020}. The prior is obtained by resampling the posterior from \cite{Tortorelli2020} using a Multivariate Gaussian. In the $\mathrm{T = 1}$ iteration, we draw $\mathrm{N = 10^4}$ samples from this prior distribution. For each set of parameters, we simulate a randomly chosen 1 square degree CFHTLS image from the W3 `Wide' field in all the 5 CFHTLS broad-bands. The number of CFHTLS W3 fields overlapping with PAUS data is $35$. At each iteration, we check that the prior and the posterior do not have too many repetitions of a single CFHTLS W3 field. This ensures that the ABC inference is not driven by the choice of a specific field. Each simulated image has the same instrumental parameters of the corresponding CFHTLS survey image and different noise seed. We then find the PAUS fields overlapping with the chosen CFHTLS image and select galaxies with $ \mathrm{i'} \le 22.5$. We simulate PAUS images for this subset of objects, using as input properties those measured from the CFHTLS images. We then match the PAUS 40 narrow-band properties with the CFHTLS ones to obtain the final catalogue on which we evaluate the distance metrics. The $43$ distance metrics in table \ref{table:tortorelli_distances} are used.

\begin{table}
\centering
\begin{tabular}{c c c}
\hline
\hline
& \textbf{Blue} & \textbf{Red} \\
\hline
$\alpha$ & -1.3 & -0.5 \\
\hline
$\mathrm{M^*_{B,slope}}$ & -0.439$^{+0.535}_{-0.652}$ & -0.697$^{+0.698}_{-0.729}$ \\
\hline
$\mathrm{M^*_{B,intcpt}} - 5 \log{h_{70}} $ & -20.623$^{+0.417}_{-0.425}$ & -20.372$^{+0.513}_{-0.466}$ \\
\hline
$\mathrm{\ln{\phi^*_{amp}}}$ / 10$^{-3}$ h$^3_{70}$ Mpc$^{-3}$ mag$^{-1}$ & -5.462$^{+0.315}_{-0.342}$ & -5.265$^{+0.602}_{-0.680}$ \\
\hline
$\mathrm{\phi^*_{exp}}$ & -0.088$^{+0.297}_{-0.277}$ & -0.836$^{+0.812}_{-0.733}$ \\
\hline
$\mathrm{r_{50,slope}^{phys}}$ & -0.242$^{+0.003}_{-0.003}$ & -0.242$^{+0.003}_{-0.003}$ \\
\hline
$\mathrm{r_{50,intcpt}^{phys}}$ & 0.871$^{+0.179}_{-0.206}$ & 0.871$^{+0.179}_{-0.206}$ \\
\hline
$\mathrm{\sigma_{phys}}$ & 0.567$^{+0.004}_{-0.004}$ & 0.567$^{+0.004}_{-0.004}$ \\
\hline
$\mathrm{a}_{1,0}$ & 1.153$^{+1.208}_{-0.766}$ & 1.293$^{+1.749}_{-0.896}$ \\ 
\hline
$\mathrm{a}_{2,0}$ & 2.209$^{+1.997}_{-1.461}$ & 1.802$^{+1.716}_{-1.179}$ \\
\hline
$\mathrm{a}_{3,0}$ & 1.589$^{+1.6187}_{-0.966}$ & 1.492$^{+1.705}_{-1.014}$ \\
\hline
$\mathrm{a}_{4,0}$ & 1.308$^{+1.484}_{-0.865}$ & 1.577$^{1.769}_{-1.001}$ \\
\hline
$\mathrm{a}_{5,0}$ & 1.908$^{+2.480}_{-1.138}$ & 1.618$^{+1.988}_{-1.079}$ \\
\hline
$\mathrm{a}_{1,1}$ & 1.287$^{+1.588}_{-0.815}$ & 1.321$^{1.702}_{-0.879}$ \\ 
\hline
$\mathrm{a}_{2,1}$ & 3.778$^{+1.820}_{-1.299}$ & 1.637$^{+1.8401}_{-0.997}$ \\
\hline
$\mathrm{a}_{3,1}$ & 0.807$^{+0.988}_{-0.497}$ & 1.558$^{+2.037}_{-1.009}$ \\
\hline
$\mathrm{a}_{4,1}$ & 1.043$^{+0.997}_{-0.613}$ & 1.404$^{+1.771}_{-0.961}$ \\
\hline
$\mathrm{a}_{5,1}$ & 1.229$^{+1.205}_{-0.686}$ & 1.703$^{+2.277}_{-1.106}$ \\
\hline
\end{tabular}
\caption{This table shows the 50th percentile values and upper and lower dispersions of the model parameters from the ABC inference on CFHTLS and PAUS survey data for red and blue galaxies. Upper and lower dispersions are expressed as the 84th - 50th percentile and the 50th - 16th percentile values, respectively.}
\label{table:gal_pop_parameters}
\end{table}

We obtain the posterior distributions at $\mathrm{T = 1}$ for all distances. We find that the distance that provides the most stringent constraints on all parameters is $\mathrm{d_{43}}$. This is the sum of: the rescaled MMD distance between magnitudes, sizes, colours and flux fractions of CFHTLS galaxies in the $\mathrm{u^{*}}$, $\mathrm{g'}$, $\mathrm{r'}$, $\mathrm{i'}$, $\mathrm{z'}$ bands, the rescaled fractional difference between detected galaxies in CFHTLS simulated and observed data, the rescaled PCA distance and the rescaled MMD distance on PCs coefficients for PAUS galaxies. This is not surprising since it is the distance that encapsulates the information on the LF parameters and size parameters through the distances measured with CFHTLS galaxies and the information on spectral coefficients through PAUS galaxies. The $\mathrm{T = 1}$ approximate Bayesian posterior consists of the $\mathrm{N_{q,1}}$ samples having distance $\mathrm{d_{43}}$ less than $\epsilon_1$, $\{ (\theta^{(\mathrm{1})}, \mathrm{d}_{43}^{(\mathrm{1})}) \} = \{ (\theta_{\mathrm{i,T=1}}, \mathrm{d}_{43,\mathrm{i,T=1}}) | \mathrm{d}_{43,\mathrm{i,T=1}} \le \epsilon_{\mathrm{1}}, \mathrm{i=1,...,N_{q,1}} \}$. The threshold $\epsilon_1$ is the $\mathrm{q}=10$-th percentile value for the distribution of the $\mathrm{d_{43}}$ distance metric.

The prior for the $\mathrm{T=2}$ iteration is obtained by resampling the approximate posterior from the $\mathrm{T=1}$ iteration with a GMM having $\mathrm{n=20}$ components. We draw $\mathrm{N'} = 10^4$ new samples and we repeat the simulation process with new randomly drawn CFHTLS and PAUS images from the W3 field. This time we only evaluate the $\mathrm{d_{43}}$ distance metric to compute the threshold $\mathrm{\epsilon_T}$. The new approximate Bayesian posterior consists of the $\mathrm{N_{q,T}}$ samples having $\mathrm{d_{43}}$ less than $\mathrm{\epsilon_T}$, $\{ (\theta^{(\mathrm{T})}, \mathrm{d}_{43}^{(\mathrm{T})}) \} = \{ (\theta_{\mathrm{i,T}}, \mathrm{d}_{43,\mathrm{i,T}}) | \mathrm{d}_{43,\mathrm{i,T}} \le \epsilon_{\mathrm{T}}, \mathrm{i=1,...,N_{q,T}} \}$. We then resample and we start the new $\mathrm{T+1}$ iteration. We keep drawing until the acceptance ratio drops below $\mathrm{p_{acc,min}} = 10\%$.

\begin{figure}[t!]
\centering
\includegraphics[width=12cm]{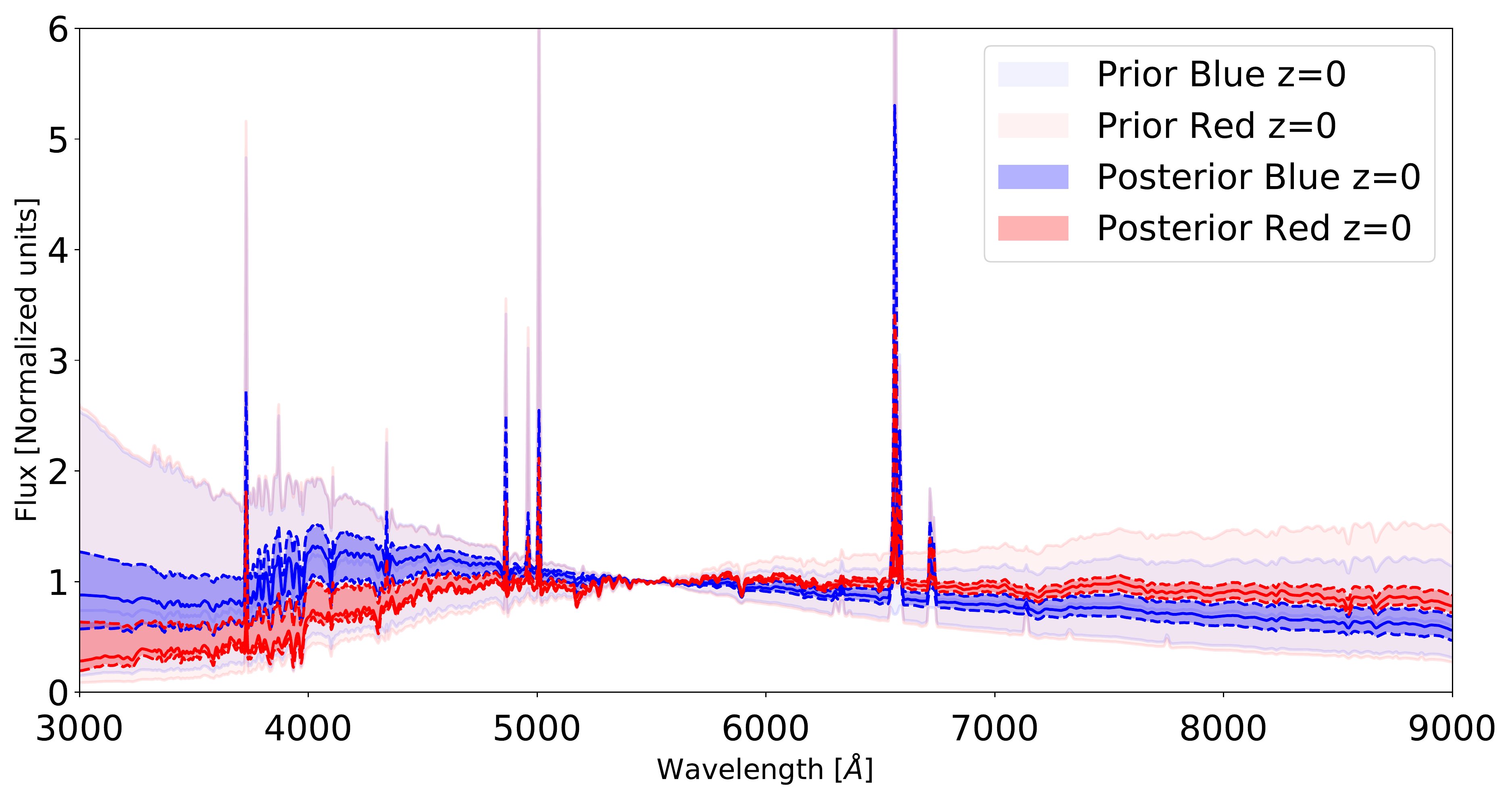}
\includegraphics[width=12cm]{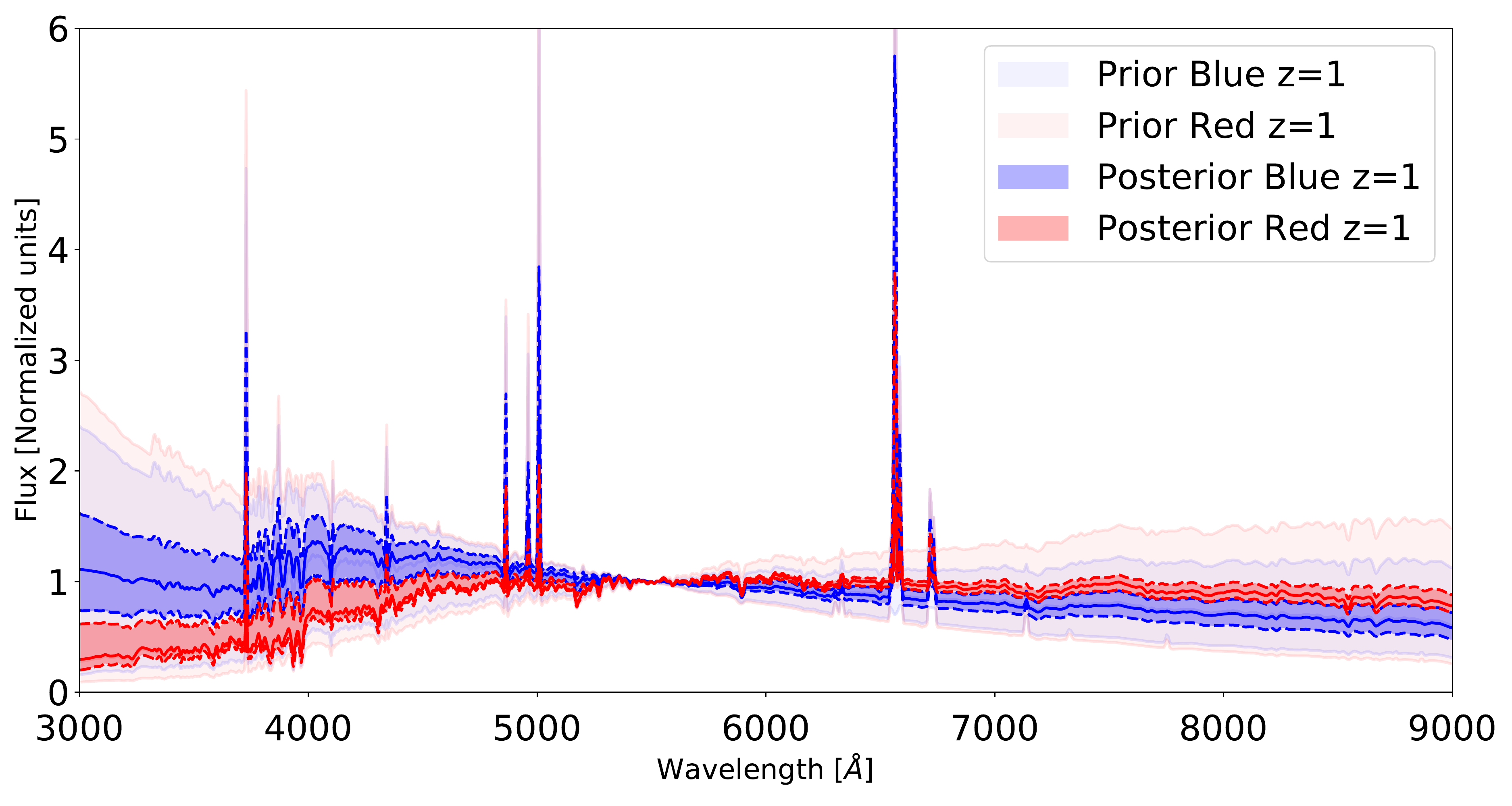}
\caption{The panels show the median and the dispersion of the simulated normalised galaxy spectra built with the spectral coefficients from the posterior distribution for $\mathrm{z=0}$ (upper panel) and $\mathrm{z=1}$ (lower panel). Each simulated spectrum is created as described in section \ref{section:gal_pop_model} using the \textsc{kcorrect} templates. Light blue and light red bands represent the dispersions of blue and red galaxies spectra from the prior distribution, while red and blue bands refer to the dispersions of galaxies from the posterior distribution. Dispersions are expressed as the 84th - 50th percentile and the 50th - 16th percentile values.} 
\label{fig:tortorelli_fig5}
\end{figure}

The acceptance ratio drops below $\mathrm{p_{acc,min}} = 10\%$ at the $\mathrm{T = 3}$ iteration. We show in figures \ref{fig:tortorelli_fig3} and \ref{fig:tortorelli_fig4} the resulting approximate Bayesian posterior distributions for the LF-size parameters and for the spectral coefficients of the galaxy population model, respectively. These results (blue contours) are over-plotted on top of the \cite{Tortorelli2020} prior (green contours) and the prior used in this work (red contours). The approximate Bayesian posterior consists of $10^3$ samples, the \cite{Tortorelli2020} prior of $10^{5}$ samples, while the prior we use in this work of $10^4$ samples. The 50th percentile values (medians) and the upper and lower dispersion of the different model parameters are reported in table \ref{table:gal_pop_parameters}. The upper and lower dispersions are expressed as the 84th - 50th percentile values and the 50th - 16th percentile values, respectively. 

In figure \ref{fig:tortorelli_fig5}, we show the median simulated spectra, and their dispersions, built using the values of the spectral coefficients $\mathrm{a_i}$ from the posterior distribution. Each simulated spectrum is created as described in section \ref{section:gal_pop_model}, meaning by multiplying the 5 \textsc{kcorrect} templates $\mathrm{T_i(\lambda)}$ for the coefficients $\mathrm{c_i}$. These are drawn from a Dirichlet distribution of order 5 parametrised by the $\mathrm{a_i}$ from the posterior distribution in figure \ref{fig:tortorelli_fig4}. Each spectrum is normalised by the median flux in the wavelength range $5450$ \AA\ $\le \lambda \le$ $5550$ \AA. The posterior distribution of the spectral coefficients correctly reproduces the differences between the blue and the red galaxy spectra. Blue galaxies have higher fluxes in the blue part of the optical spectrum ($\lambda \lesssim 4500$ \AA) and smaller fluxes in the red part ($\lambda \gtrsim 7000$ \AA) with respect to red galaxies. Moreover, blue galaxies are characterised by stronger emission lines than red galaxies. The evolution of the red galaxies from redshift $\mathrm{z=1}$ to $\mathrm{z=0}$ is shallower than that of blue galaxies, which have higher fluxes in the blue part of the spectrum and stronger emission lines at higher redshift. This picture is coherent with red galaxies being characterised by an older stellar population than blue galaxies and with literature studies that find that red galaxies evolve little from $\mathrm{z=1}$ to $\mathrm{z=0}$, while blue galaxies are chacterised by a higher level of star-formation at higher redshifts (e.g., \cite{Brown2007,Madau2014,Tortorelli2020}).

The addition of PAUS data and its contribution in the distance metrics help to break the degeneracy in the spectral coefficients with respect to the ABC inference in \cite{Tortorelli2020}. Since the information provided by CFHTLS data has already been mostly saturated in \cite{Tortorelli2020}, the addition of the PAUS dataset provides only marginal gain to the LF and size constraints. To quantify that, we use the definition for a relative generalised figure of merit presented in \cite{Wang2008}. We compute the covariance matrices of our 31 parameter space for the posterior distribution in \cite{Tortorelli2020} and this work. The figure of merit is then the inverse of the square root of the covariance matrix determinant. From the ratio of the figures of merit, we find that adding PAUS data improves its value by a factor of $\sim27$ with respect to CFHTLS data alone.

\section{Principal Component Analysis on Narrow-band Spectra}
\label{section:pca_nb_spectra}

This section presents the results from the PCA performed on observed and simulated galaxies drawn from the approximate posterior distribution. The aim of this section is to show that the constrained spectral coefficients of the galaxy population model are able to output simulated galaxies that show the same spectral characteristics as the real ones. To do that, we first perform the PCA in the observed-frame and then in the rest-frame. The observed-frame PCA uses the fluxes directly measured on narrow-band images. The rest-frame PCA, instead, uses the rest-frame luminosities obtained by performing an SED fitting with \textsc{CIGALE} (see section \ref{section:stellar_pop}) \cite{Burgarella2005,Noll2009,Boquien2019} on a sub-sample of galaxies with magnitude $ 18 \le \mathrm{i'} \le 22.5$. The observed-frame PCA focuses on the comparison of the galaxy spectral shapes as directly measured and therefore it serves us as a test to see whether we are able to properly simulate real fluxes and the same observational effects that appear on observed data and simulations. The rest-frame PCA, being based on luminosities from SED fitting, focuses on a more physical comparison of the galaxy spectral features, testing whether the individual spectral features are the same between observations and simulations.

\subsection{Observed-frame Principal Component Analysis}
\label{subsection:observed_frame_pca}

\begin{figure}[t!]
\centering
\includegraphics[width=7.65cm]{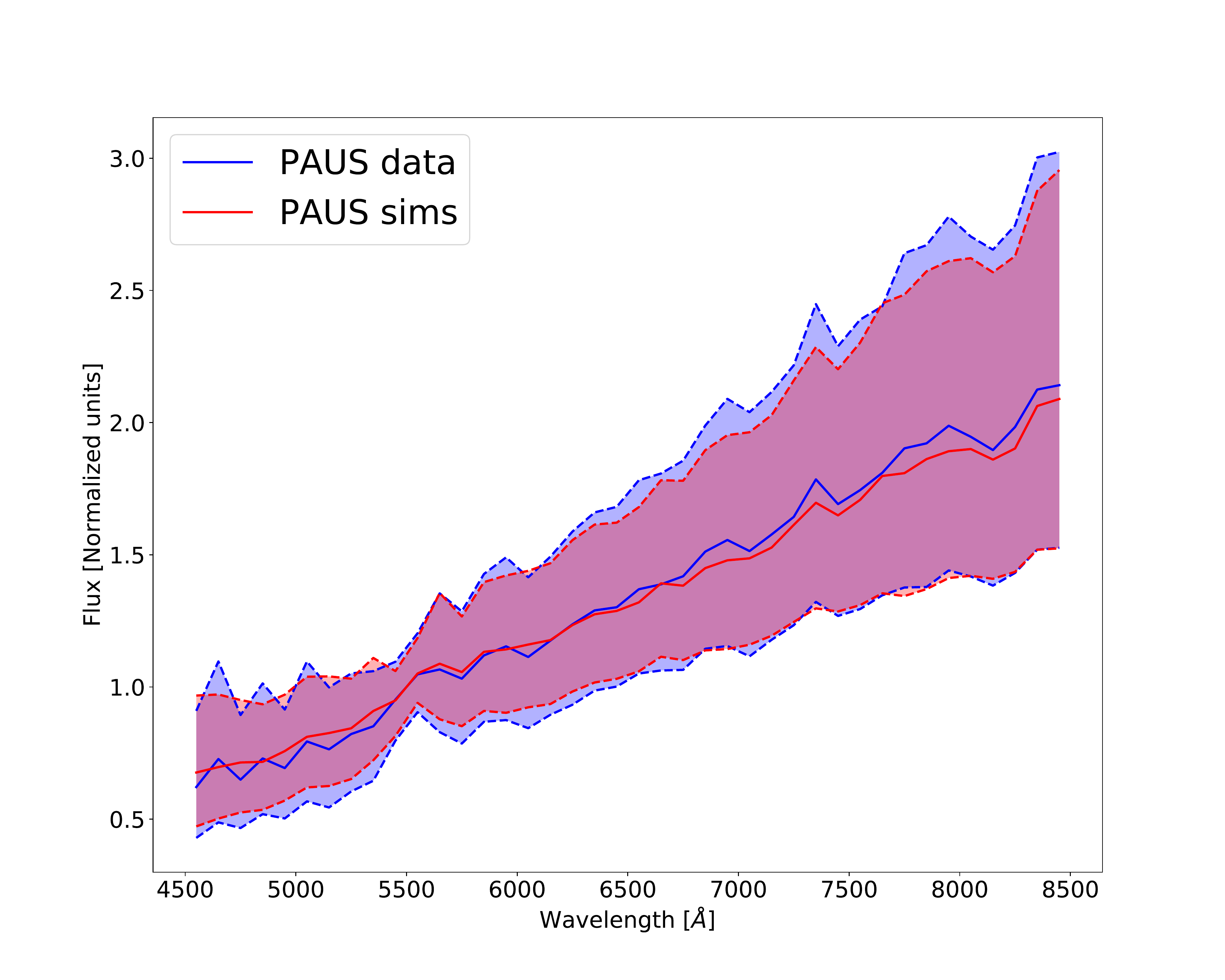}
\includegraphics[width=7.65cm]{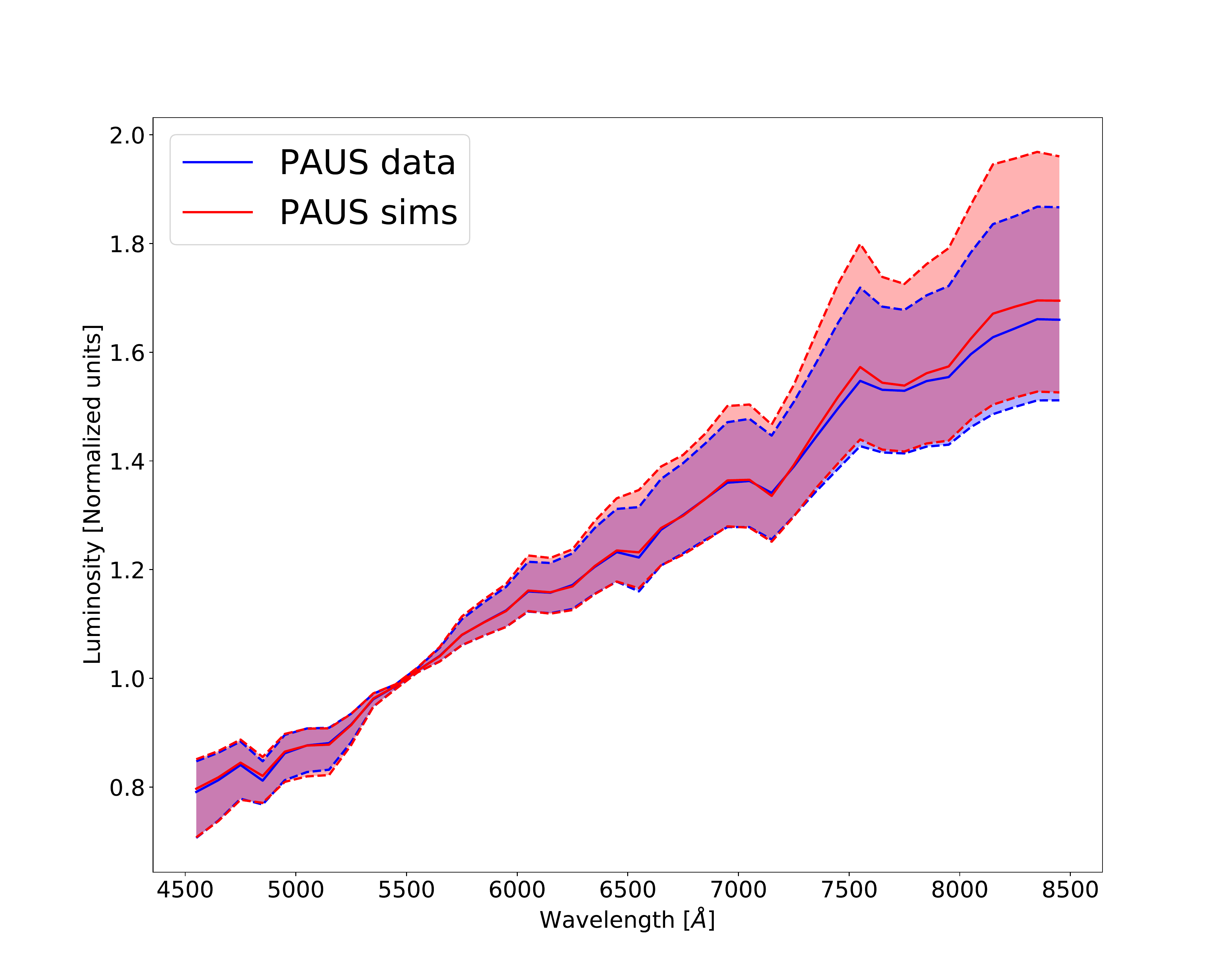}
\caption{The left panel shows the comparison between the observed-frame median spectra of PAUS simulated (red) and observed (blue) narrow-band data. The right panel shows the same comparison, but in the rest-frame. The spectra have been normalised by the median flux in the wavelength range $5450$ \AA\ $\le \lambda \le$ $5550$ \AA. The red and blue lines show the median normalised flux in the 40 narrow-bands, while the dashed lines show the 25th and 75th percentile values of normalised fluxes.} 
\label{fig:tortorelli_fig6}
\end{figure}

We first compare the median observed-frame narrow-band spectra of observations and simulations. To build them, we use the PAUS fluxes of observed and simulated galaxies drawn from the posterior distribution. We recall that these galaxies have been selected as those objects with $\mathrm{i'} \le 22.5$ in the CFHTLS images (see section \ref{subsection:paus_sims_analysis}) and that the fluxes have been calibrated to have the same zero-point in all wavebands.

To build the median spectra, we first normalise the PAUS fluxes of each galaxy by its median flux in the wavelength range $5450$ \AA\ $\le \lambda \le$ $5550$ \AA.\footnote{PAUS observed data have a mean signal-to-noise of $\mathrm{S/N \sim 3}$ in this wavelength range. The highest mean signal-to-noise is roughly $\mathrm{S/N \sim 5}$ in the reddest bands.} This normalisation preserves the overall shape of the spectrum. The choice of the wavelength range for the normalisation is not trivial. We avoid the use of regions where the filter throughputs are low ($\lambda \lesssim 5000$ \AA) and regions where the sky contribution is stronger ($\lambda \gtrsim 7000$ \AA). We choose the range $5450$ \AA\ $\le \lambda \le$ $5550$ because it lacks strong emission lines at the median redshift ($\mathrm{z \sim 0.4}$) of our sample. Then, we take the quartile flux values in each wavelength bin for all galaxies. We show in the left panel of figure \ref{fig:tortorelli_fig6} the median spectra for PAUS observed galaxies (blue line and bands) and PAUS simulated galaxies (red line and bands). The agreement of the median spectra is very good, both in terms of the overall spectral shape (red and blue lines) and of the flux measurements dispersion (red and blue bands). This is also confirmed by the value of the mean difference between the normalised spectral shapes, $\Delta \mathrm{f} = 0.02 \pm 0.04$, that is consistent with zero within errors. Since in the observed-frame the individual spectral features are not visible, this comparison is particularly informative on the variance of the narrow-band spectra that is consistent between the two datasets.

\begin{figure}
\centering
\includegraphics[width=7.65cm]{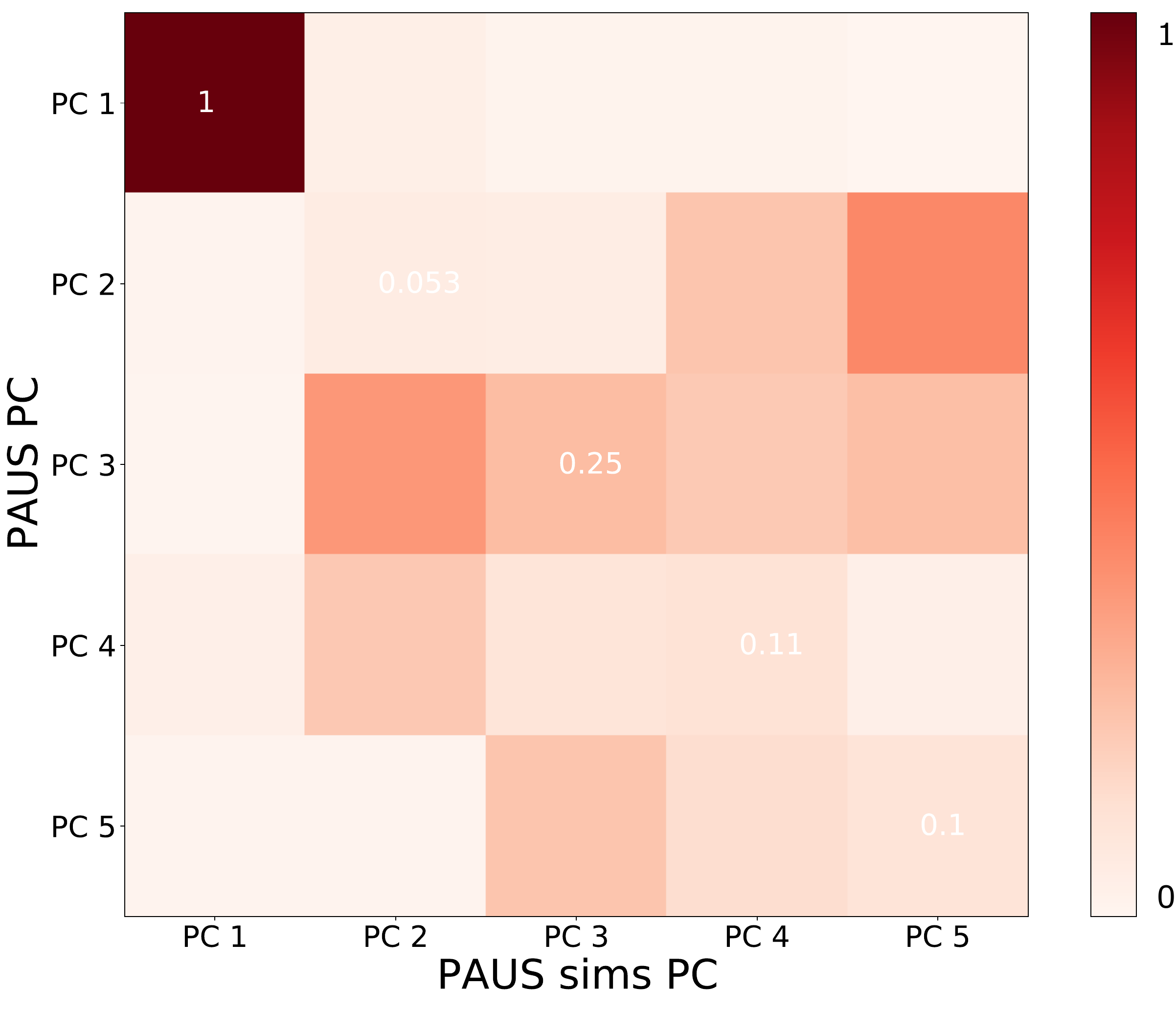}
\includegraphics[width=7.65cm]{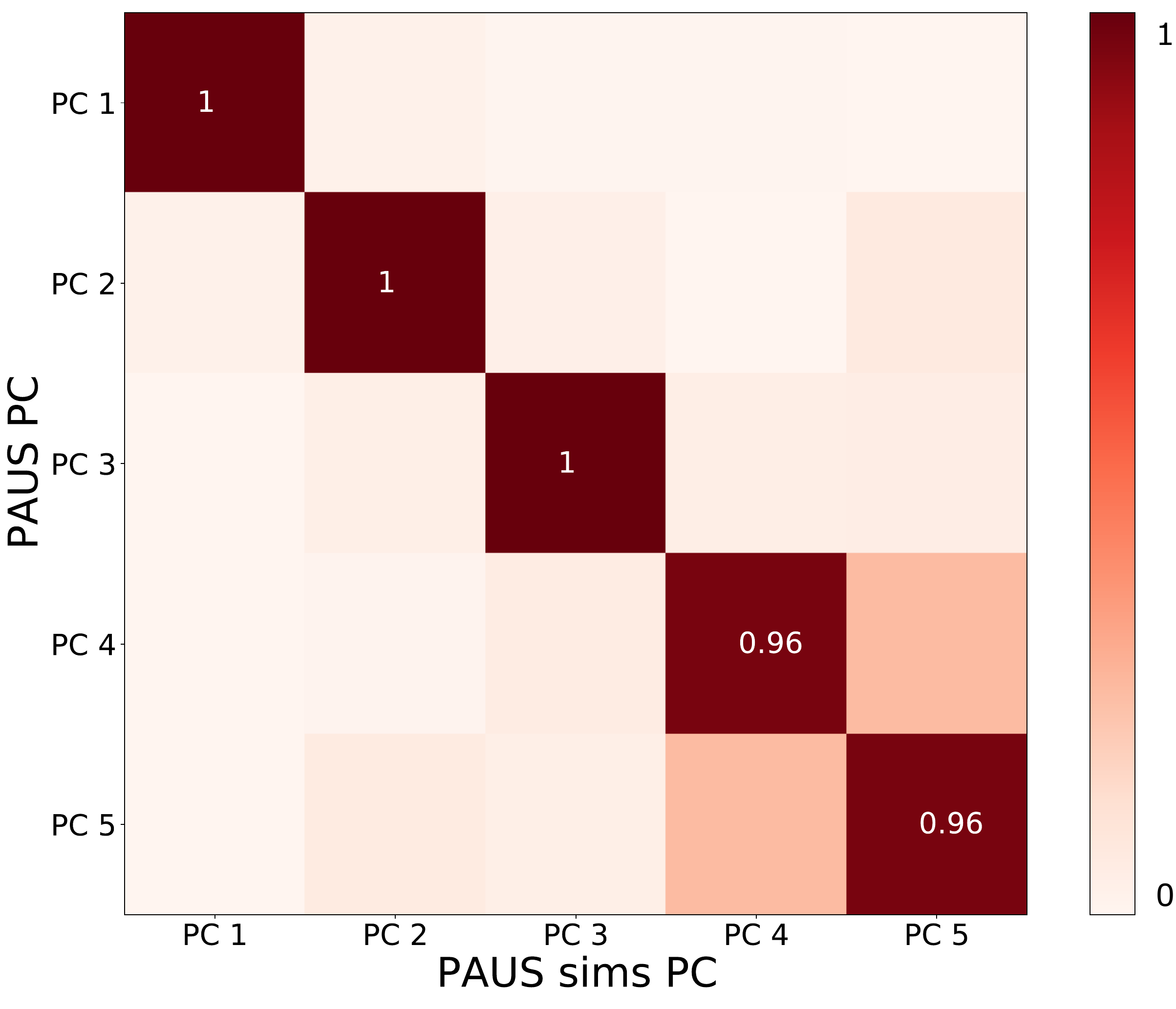}
\caption{The left panel shows the mixing matrix between the observed-frame observed and simulated PAUS PCs. The right panel shows the matrix for the rest-frame PAUS PCs. The horizontal axes refer to the simulated PCs, while the vertical axes to the observed ones. Being orthonormal, the colour range goes from 0 to 1.}
\label{fig:tortorelli_fig7}
\end{figure}

\begin{figure}
\centering
\includegraphics[width=16cm]{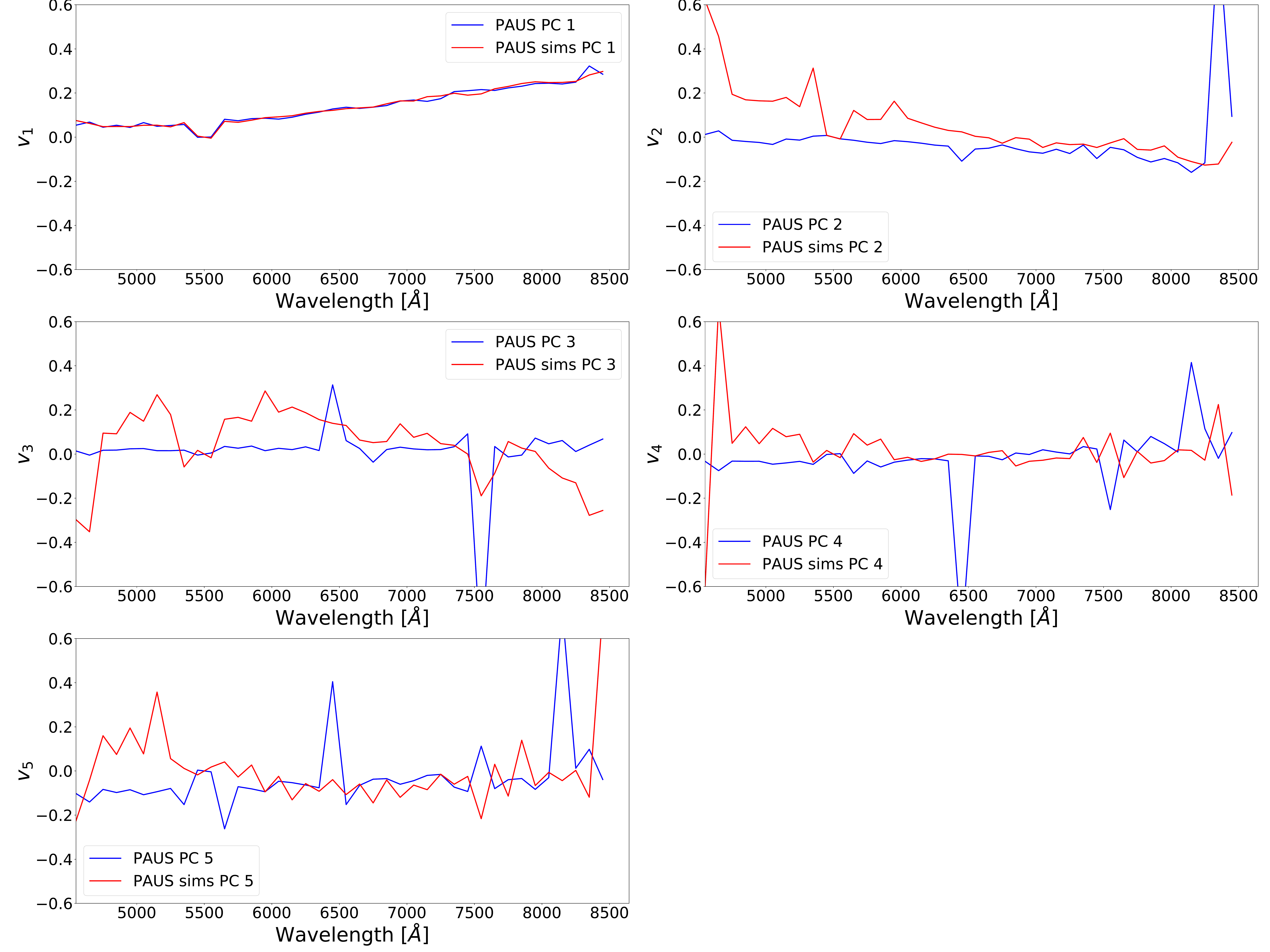}
\caption{From top left to bottom left panels, we show the first 5 observed and simulated PCs resulting from the PCA in the observed-frame. Blue lines represent observed PAUS PCs, while red lines represent simulated ones. The horizontal axes show the wavelength in \r{A}ngstrom, while the vertical axes show the normalised PC value.}
\label{fig:tortorelli_fig8}
\end{figure}

PAUS data contains more spectral information than a broad-band dataset. Comparing observed data and simulations on the basis of colours among 40 different narrow-bands is therefore not an efficient option, since it does not allow us to immediately visualise the information about the overall spectral shape. To capture the global information from such a high dimensional dataset, we follow the approach we already adopted in \cite{Tortorelli2018b}, meaning we perform a PCA on the observed-frame PAUS narrow-band spectra. We perform the PCA following the same steps highlighted in section \ref{subsection:distance_metrics}.

The left panel of figure \ref{fig:tortorelli_fig7} shows the resulting mixing matrix obtained by performing the scalar product between the observed and simulated PCs. The matrix shows an excellent agreement of the first PC ($\phi_1 \cdot \psi_1 = 1.00$), while the noise is dominating from the second PC on.  The actual PCs are shown in figure \ref{fig:tortorelli_fig8}. Since the analysis is performed in the observed-frame, the same spectral feature in galaxies at different redshifts occurs at different wavelengths. This causes a smoothing of the spectra resulting from the PCA. Therefore, the first PC is only able to capture the overall spectral shape of the galaxy population, without any visible spectral feature. Furthermore, this causes the higher order PCs to be mostly noise dominated, such that no dominant spectral feature is visible in them as well. %Part of the discrepancy of high order components may also be addressed to the different redshift estimations of the observed data and simulations. While the latter have exact redshifts that are drawn from the input LFs, the former redshifts are photo-z estimates obtained with the BCNZ2 code \cite{Alarcon2021}. This may also explain why some signal is present in the higher order PCs of the simulations, but not of the observed data.

\begin{figure}
\centering
\includegraphics[width=6.90cm]{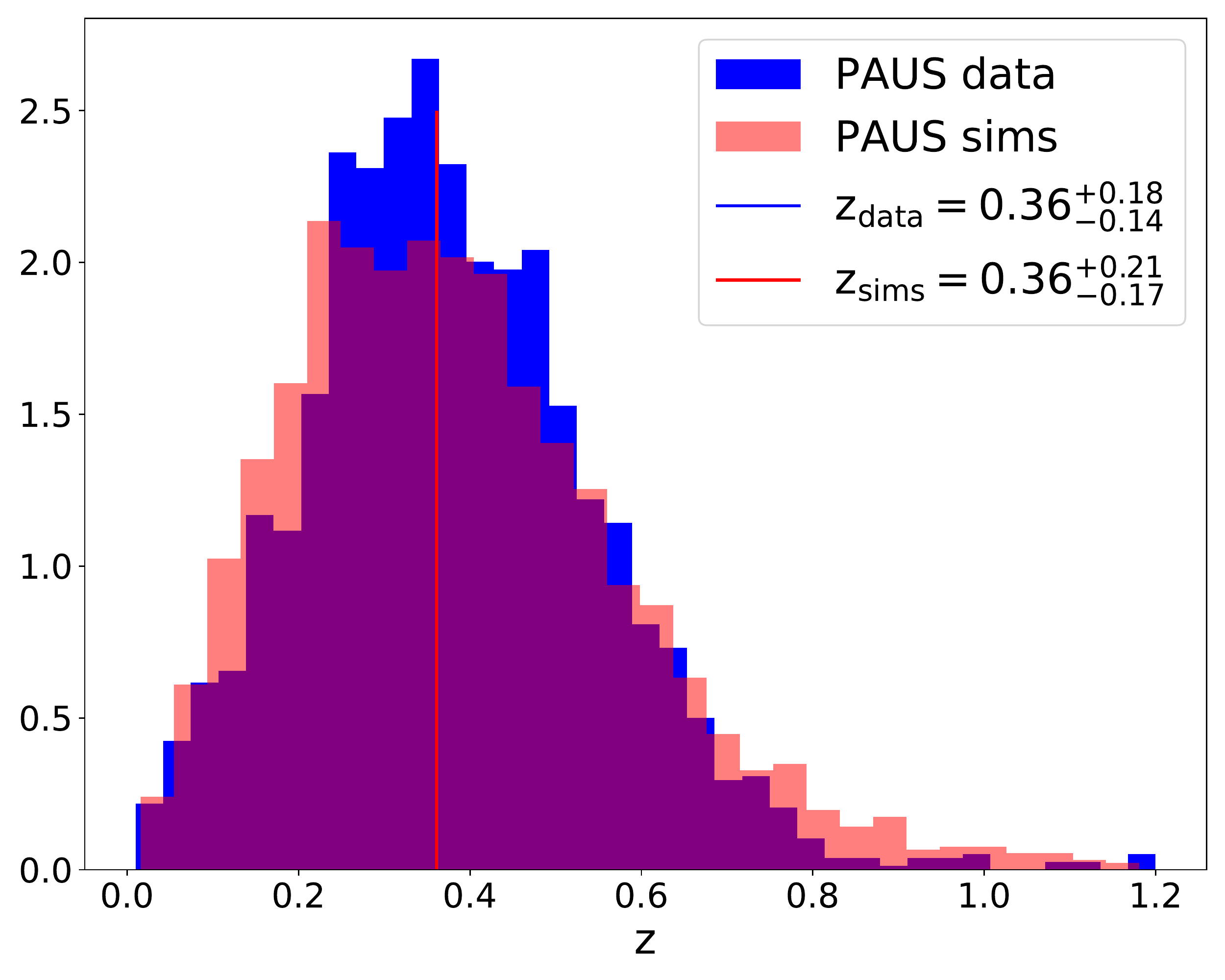}
\includegraphics[width=7.65cm]{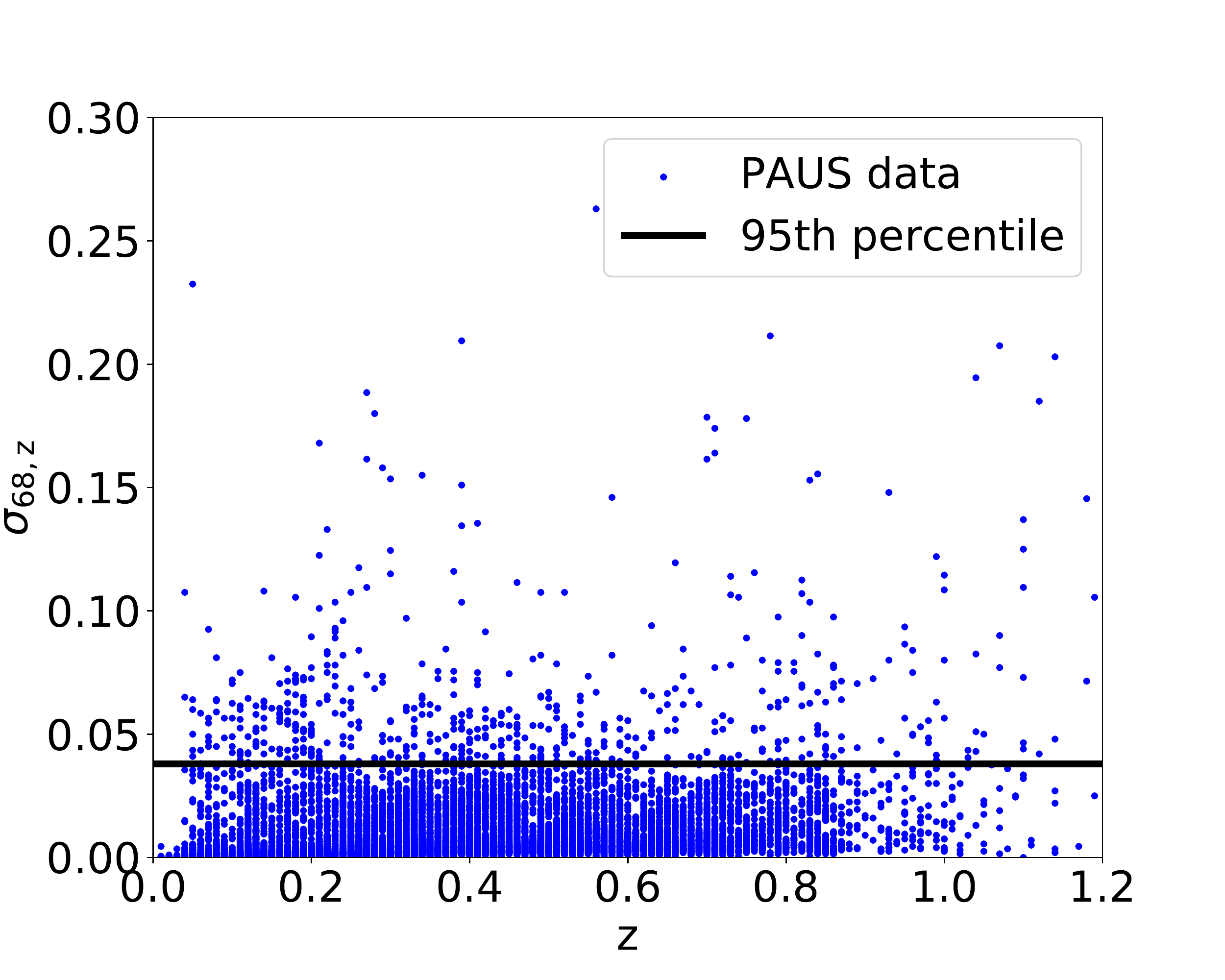}
\caption{The left panel shows the comparison between the redshift distributions of observed (blue histogram) and simulated galaxies (red histogram). The sub-sample of objects having $18 \le i^{'} \le 22.5$ is shown. We report also the median and upper and lower dispersion. They are expressed as the 84th - 50th percentile values and the 50th - 16th percentile values, respectively. Redshifts from simulation have no errors as they are drawn from the galaxy population model. Redshifts from observations, instead, are estimated using the BCNZ2 code \cite{Alarcon2021}. The right panel shows the errors on redshift, expressed as the $\sigma_{68,z}$, as function of the redshift itself. $95\%$ of our flux-limited sample of observed galaxies have uncertainties smaller than 0.038.}
\label{fig:tortorelli_fig9}
\end{figure}

\begin{figure}
\centering
\includegraphics[width=16cm]{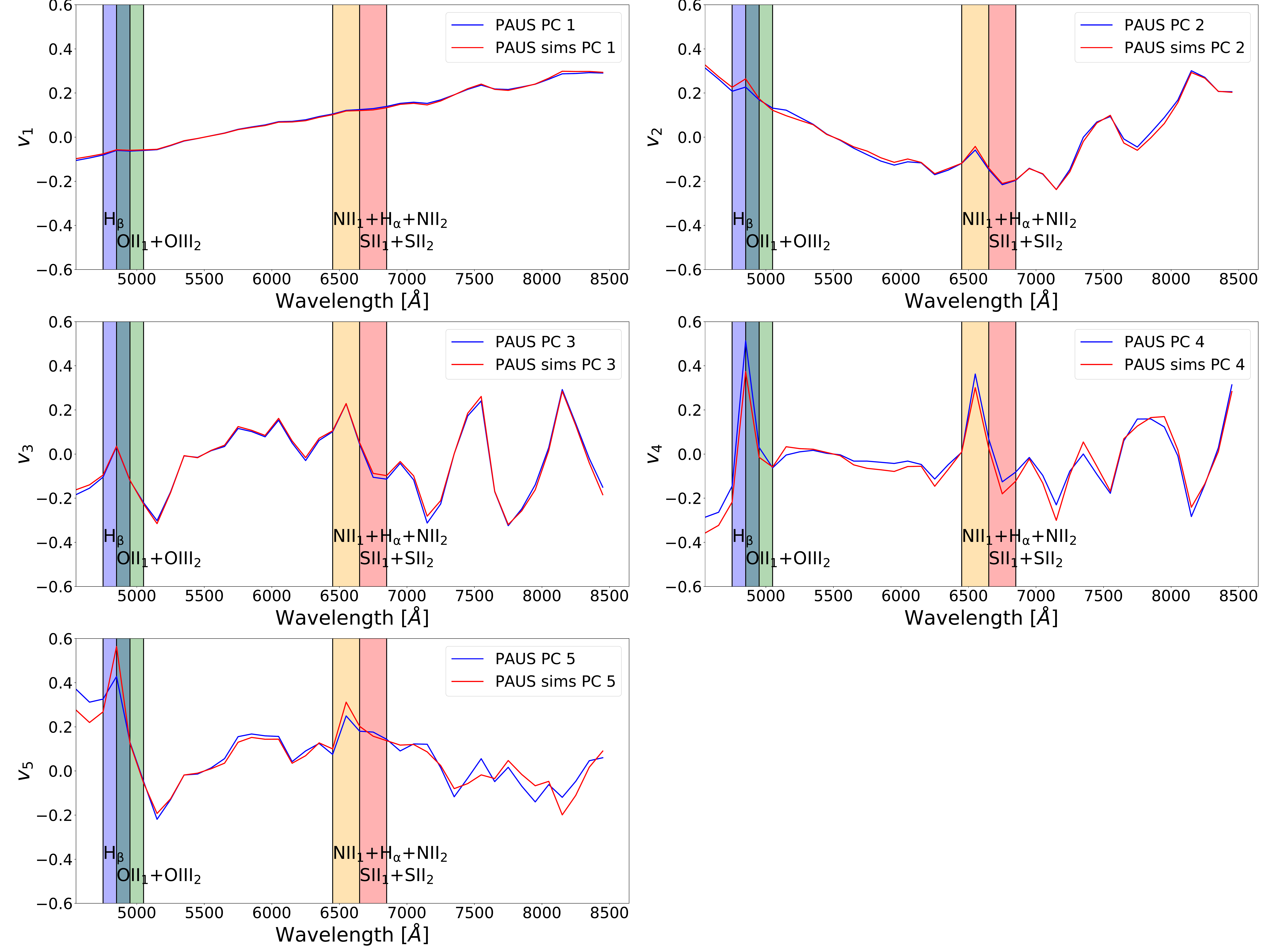}
\caption{From top left to bottom left panels, we show the first 5 observed and simulated PCs resulting from the PCA in the rest-frame. Blue lines represent observed PAUS PCs, while red lines represent simulated ones. The coloured vertical bands represent the narrow-band range where common emission lines fall. The light blue, green, orange and red vertical bands refer to the $\mathrm{H_{\beta}}$, $\mathrm{OII_{1} + OIII_{2}}$ complex, $\mathrm{NII_{1} + H_{\alpha} + NII_{2}}$ complex and $\mathrm{SII_{1} + SII_{2}}$ complex, respectively. The horizontal axes show the wavelength in \r{A}ngstrom, while the vertical axes show the normalised PC value.}
\label{fig:tortorelli_fig10}
\end{figure}

\subsection{Rest-frame Principal Component Analysis}
\label{subsection:rest_frame_pca}

The PCA in the observed-frame is already able to capture the bulk of the spectral information from the PAUS dataset. However, spectral features, e.g., emission lines, are broadened by the different redshifts at which they appear in different galaxies. To capture the full spectral diversity of PAUS data, we need to perform the PCA in the rest-frame.

The sample of galaxies on which we perform the rest-frame PCA is constituted by objects drawn from the posterior distribution having magnitude in the range $ 18 \le \mathrm{i'} \le 22.5$. These are $11031$ and their matching redshift distributions are shown in figure \ref{fig:tortorelli_fig9}. The observed galaxy redshifts are obtained by matching them with the catalogue used in \cite{Johnston2021}. Simulated galaxies have redshifts drawn from the galaxy population model.

We use the rest-frame luminosities obtained by performing an SED fitting with the \textsc{CIGALE} \cite{Burgarella2005,Noll2009,Boquien2019} code on the observed and the simulated galaxies. The SED fit combines the information from broad-band and narrow-band photometry (see section \ref{section:stellar_pop} for details).  \textsc{CIGALE} outputs the rest-frame luminosities in the 40 narrow-bands for each observed and simulated galaxy. These rest-frame luminosities are computed from the best-fitting spectra integrated in the PAUS narrow-bands. The right panel of figure \ref{fig:tortorelli_fig6} shows the median rest-frame narrow-band spectra for PAUS observed galaxies (blue line and bands) and PAUS simulated galaxies (red line and bands). Up to $\lambda \sim 7500$ \AA, the agreement of the median values is very good. From $\lambda \gtrsim 7500$ \AA, the median values are slightly discrepant although within the dispersion.

As described in \ref{subsection:observed_frame_pca}, we perform the PCA in the rest-frame with $\mathrm{n=5}$ components using the mean subtracted normalised narrow-band spectra. We show in the right panel of figure \ref{fig:tortorelli_fig7} the resulting rest-frame mixing matrix. The matrix shows an excellent agreement between observed and simulated PCs up to the 5th PC. The actual PCs are instead shown in figure \ref{fig:tortorelli_fig10}. The first PC captures the red continuum of the galaxy population, the second component the blue continuum due to star-formation and from the third component on spectral features due to emission lines are visible (coloured vertical bands). The slight differences in the fourth and fifth PCs may be due to the slightly different strengths of emission lines between observed data and simulations and to the different redshift estimation methods. In particular, the redshifts of simulated galaxies have no errors since they are drawn from the galaxy population model, while the uncertainties on observed data are estimated using the photo-z code BCNZ2 \cite{Alarcon2021}. Despite that, the effect of the different estimations is minor, since the redshift distributions are in good agreement between observed data and simulations (see right panel of figure \ref{fig:tortorelli_fig9}). Furthermore, $95\%$ of our flux-limited sample of observed galaxies have errors on the photo-z estimate, expressed as $\sigma_{68,z}$, lower than 0.038.

The observed-frame and rest-frame PCA show that the region of the parameter space we constrain with ABC correctly captures the diversity of the spectral characteristics visible in the PAUS data. This proves that our methodology is able to exploit the full potential of this rich dataset. Furthermore, it proves that the galaxies drawn from simulations replicate the spectral features visible in the analysed flux-limited galaxy population.

\section{Stellar population properties}
\label{section:stellar_pop}

In this section, we further test the quality of our results and the ability of our constrained galaxy population model to create simulated galaxies with realistic stellar population properties. We measure the stellar population properties of observed and simulated galaxies through SED fitting. We use the code \textsc{CIGALE} \cite{Burgarella2005,Noll2009,Boquien2019} to measure galaxy properties. \textsc{CIGALE} has been developed to find the best-fit galaxy SED by comparing galaxy spectral models with observations spanning the wavelength range from X-ray to far-infrared and radio. \textsc{CIGALE} estimates the physical properties of galaxies by evaluating a large grid of models on data to minimise the likelihood distribution. The model SED relies on Bruzual and Charlot stellar evolution synthesis models \cite{bruzual03} that are combined with flexible star-formation histories (hereafter, SFHs) to create the initial galaxy SED. \textsc{CIGALE} then adds emission from gas ionised by massive stars, it attenuates the galaxy light on the base of a flexible attenuation curve and it adds the contribution of far-infrared emission coming from energy reprocessed by dust. The particular setup we use to fit our galaxy SED is the same used in \cite{Johnston2021}. We use Bruzual and Charlot models \cite{bruzual03}, a SFH scenario modelled as a delayed SFR with an optional exponential burst, modified Calzetti attenuation law \cite{Calzetti2000}, Draine \cite{Draine2014} dust emission model and Chabrier IMF \cite{Chabrier2003}. The stellar templates in \textsc{CIGALE} are the same that are used to create the \textsc{kcorrect} templates we build our simulated spectra with. Gas emission line physics, instead, is modelled in \textsc{CIGALE} using \textsc{CLOUDY} \cite{Ferland1998}, while in our simulations it is based on the MAPPINGS-III \cite{Allen2008} templates.

The observed and the simulated galaxies are the same ones used in section \ref{subsection:rest_frame_pca}, spanning the redshift range from $\mathrm{z=0}$ to $\mathrm{z \sim 0.8}$. We perform the SED fitting with \textsc{CIGALE} using both the 5 broad-bands from CFHTLS and the 40 narrow-bands from PAUS. The contemporary fit of broad-bands and narrow-bands is carried out for two different setups of \textsc{CIGALE}. In the first, we fix the metallicity to the solar value $\mathrm{\log{(Z/Z_{\odot})}}=0$. In the second setup, we let the metallicity free to vary in the range $-2.3 \le \log{(Z/Z_{\odot})} \le 0.4$.

The observed and simulated photometric SEDs are well fitted by \textsc{CIGALE}. The quality of the fits is quantified using the reduced chi-square $\chi^2_{\mathrm{r}}$. In \textsc{CIGALE}, the $\chi^2_{\mathrm{r}}$ is not exactly equal to its statistical definition. Since the models are highly non-linear, the number of degrees of freedom for each galaxy is unknown and it is extremely difficult to quantify this number properly \cite{Andrae2010,Chevallard2016}. Therefore, we follow the approach highlighted in \cite{Malek2014,Malek2017,Malek2018} for both observations and simulations. We remove bad fits by sigma-clipping the $\chi^2_{\mathrm{r}}$ value distribution and keep only those galaxies for which the reduced chi-square $\chi^2_{\mathrm{r}} < \bar{\chi^2_{\mathrm{r}}} + 3\sigma$, where $\bar{\chi^2_{\mathrm{r}}}$ is the mean chi-square and $\sigma$ is its standard deviation. The effect of this selection is discussed in appendix \ref{appendix:uvj_selection}. This selection does not modify the shape and the median value of the redshift distribution in figure \ref{fig:tortorelli_fig9}. After this selection, we are left with $7101$ observed and $6236$ simulated galaxies, when fitting with fixed solar metallicity, whereas with 7135 observed and 6260 simulated galaxies, when we let the metallicity free to vary. The results reported in section \ref{section:sed_solar}, section \ref{section:sed_variable} and in table \ref{table:stellar_pop_values} are based on the sub-sample of galaxies with good quality of the fit. 

To validate the reliability of our simulations against different galaxy types, we further split galaxies into quiescent (QS) and star-forming (SF) using the UVJ diagram \cite{Wuyts2007,Williams2009} (see appendix \ref{appendix:uvj_selection}). This has already been proven to be effective in separating red and blue galaxies for our simulations, as discussed in appendix D of \cite{Tortorelli2020}.

\begin{table}
\centering
\resizebox{\textwidth}{!}{
\begin{tabular}{p{1.4cm} p{2.6cm} p{2.6cm} p{2.8cm} p{2.8cm}}
\hline
\hline
& \textbf{Obs $\mathrm{Z_{\odot}}$} & \textbf{Sims $\mathrm{Z_{\odot}}$} & \textbf{Obs $\mathrm{Z_{var}}$} & \textbf{Sims $\mathrm{Z_{var}}$}\\
\hline
$A_{V}$ [mag] & QS: 0.53$\pm$0.25 \newline SF: 0.73$\pm$0.26& QS: 0.43$\pm$0.25 \newline SF: 0.69$\pm$0.28& QS: 0.56$\pm$0.26 \newline SF: 0.63$\pm$0.30& QS: 0.52$\pm$0.25\newline SF: 0.64$\pm$0.30\\
\hline
$\tau$ \quad [Gyr]  & QS: 0.10$\pm$0.14 \newline SF: 1.60$\pm$1.78& QS: 0.20$\pm$ 0.21 \newline SF: 1.47$\pm$1.72& QS: 0.22$\pm$0.27 \newline SF: 1.45$\pm$1.74 & QS: 0.26$\pm$0.29 \newline SF: 1.29$\pm$1.64\\
\hline
$\mathrm{Age_{M_{*}}}$ [Gyr] & QS: 1.55$\pm$0.72 \newline SF: 1.32$\pm$1.20& QS: 2.72$\pm$1.12 \newline SF: 2.48$\pm$1.67& QS: 3.42$\pm$1.90 \newline SF: 2.02$\pm$1.76& QS: 4.38$\pm$2.01 \newline SF: 3.27$\pm$2.00 \\
\hline
$\mathrm{\log{(Z/Z_{\odot})}}$ & QS: 0.00 \newline SF: 0.00& QS: 0.00 \newline SF: 0.00& QS: -0.35$\pm$0.26 \newline SF: -0.37$\pm$0.20& QS: -0.25$\pm$0.24 \newline SF: -0.17$\pm$0.11\\
\hline
$\mathrm{\log{(SFR)}}$ [$\mathrm{M_{\odot} / yr}$] & QS: -0.42$\pm$0.30 \newline SF: 0.66$\pm$0.25& QS: -0.33$\pm$0.27 \newline SF: 0.70$\pm$0.25& QS: -0.36$\pm$0.33 \newline SF: 0.46$\pm$0.36& QS: -0.30$\pm$0.29 \newline SF: 0.66$\pm$0.29\\
\hline
$\mathrm{\log{(M_{*})}}$ [$\mathrm{M_{\odot}}$] & QS: 10.35$\pm$0.12 \newline SF: 9.76$\pm$0.17& QS: 10.48$\pm$0.12 \newline SF: 9.91$\pm$0.18& QS: 10.42$\pm$0.16 \newline SF: 9.76$\pm$0.19& QS: 10.50$\pm$0.15 \newline SF: 9.95$\pm$0.19\\
\hline
\end{tabular}
}
\caption{This table shows the median values and the median errors of the stellar population parameters estimated with \textsc{CIGALE}. The quoted parameters are obtained from the property distributions of galaxies with reduced chi-square $\chi^2_{\mathrm{r}} < \bar{\chi^2_{\mathrm{r}}} + 3\sigma$. `QS' stands for quiescent, while `SF' for star-forming. `obs' and `sims' are short for observations and simulations. The physical units of each parameter are reported in the table.}
\label{table:stellar_pop_values}
\end{table} 

\subsection{SED fitting with solar metallicity}
\label{section:sed_solar}

\begin{figure}[t!]
\centering
\includegraphics[width=12cm]{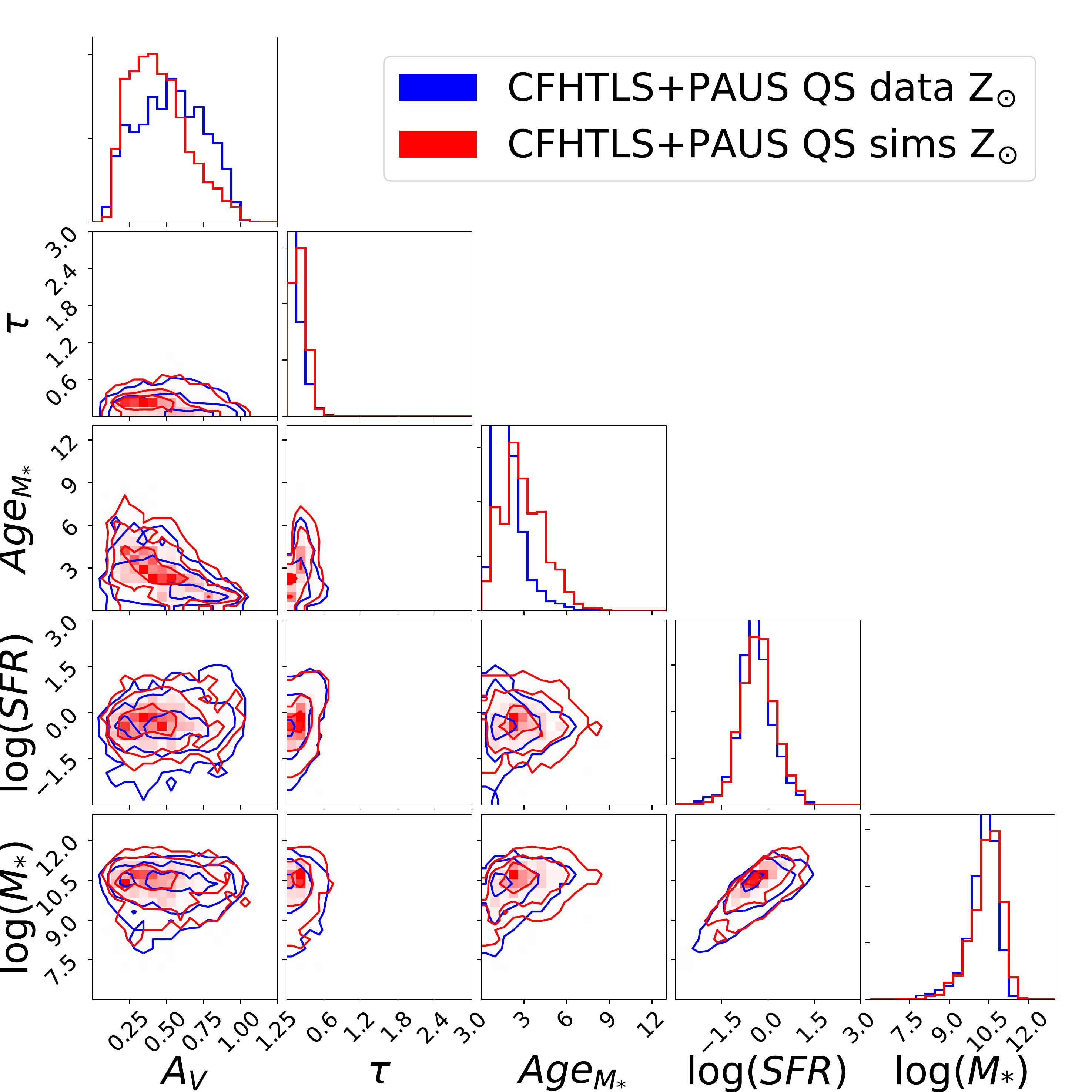}
\caption{This figure shows the comparison between the stellar population properties of QS observed (blue) and simulated (red) galaxies. The properties are measured using the SED fitting code \textsc{CIGALE} with solar metallicity. The fit uses the broad-band photometric information from CFHTLS and narrow-band photometric information from PAUS. $A_V$ is in units of mag, $\tau$ and $\mathrm{Age_{M_{*}}}$ in units of Gyr, SFR in units of $\mathrm{M_{\odot} / yr}$, while $\mathrm{\log{(M_{*})}}$ is in units of $\mathrm{M_{\odot}}$.}
\label{fig:tortorelli_fig11}
\end{figure}

\begin{figure}[t!]
\centering
\includegraphics[width=12cm]{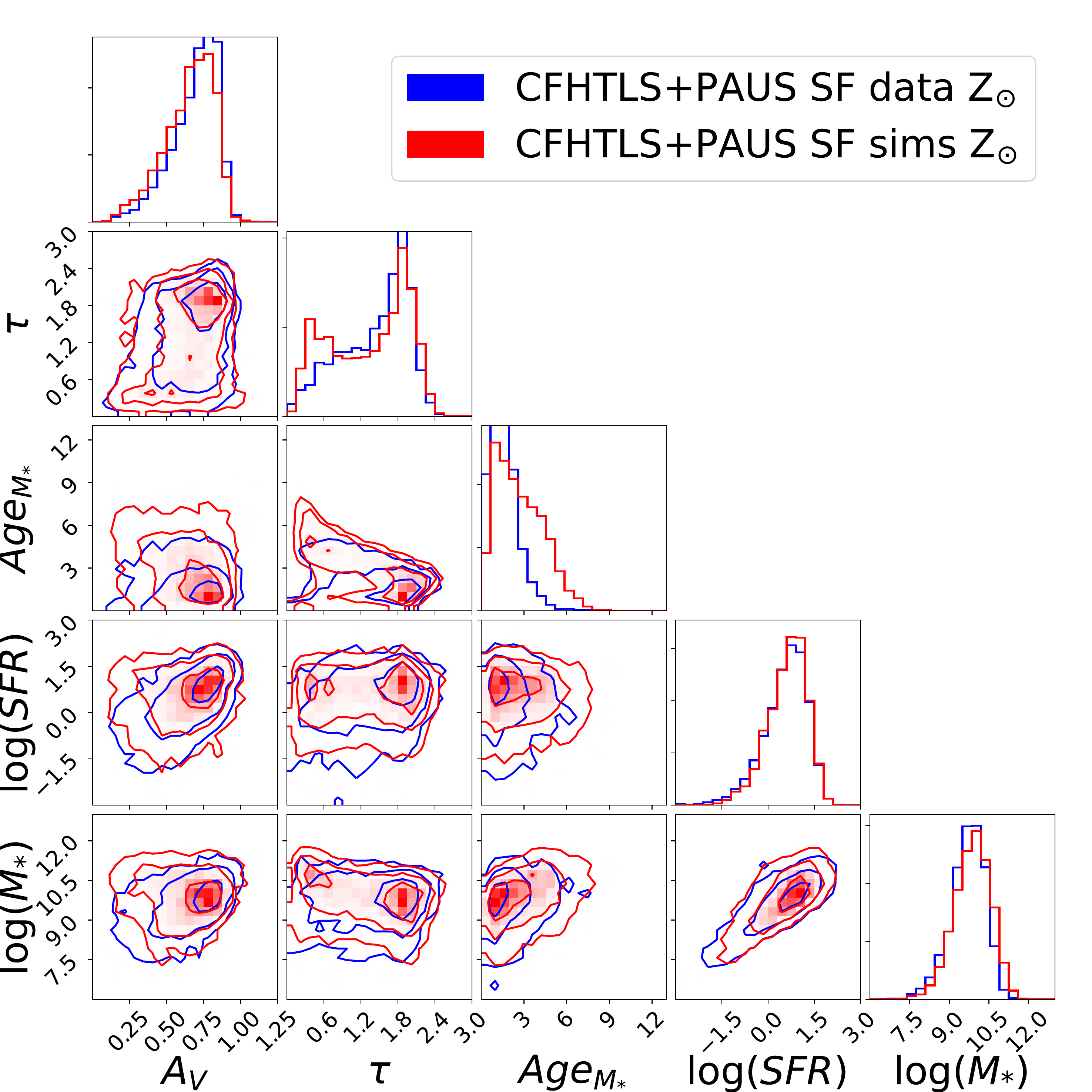}
\caption{This plot shows the same comparison as in figure \ref{fig:tortorelli_fig11}, but for SF galaxies.}
\label{fig:tortorelli_fig12}
\end{figure}

Figures \ref{fig:tortorelli_fig11} and \ref{fig:tortorelli_fig12} show the comparison between the stellar population parameters we estimate on observations and simulations for QS and SF galaxies, respectively. The SED fitting has been performed by simultaneously fitting the broad-bands and the narrow-bands and by keeping the metallicity fixed at its solar value, $\mathrm{\log{(Z/Z_{\odot})}}=0$. The measured properties are the attenuation in the V-band $A_V$, the e-folding time $\tau$ of the SFH delayed component, the mass-weighted stellar age $\mathrm{Age_{M_{*}}}$, the integrated SFR $\mathrm{\log{(SFR)}}$ and the stellar-mass $\mathrm{\log{(M_{*})}}$. Median values and median errors for these parameters are reported in table \ref{table:stellar_pop_values}. 

Our model is able to mirror adequately the observed main galaxy population, as QS and SF galaxies are clearly separated. QS galaxies, identified through the UVJ diagram, are typically more massive, with lower SFRs, shorter e-folding times and older stellar populations than SF galaxies. Moreover, we find that the stellar masses have a bimodal distribution, peaking at around $\mathrm{\log{(M_{*})}} \sim 10.4$ and $\mathrm{\log{(M_{*})}} \sim 9.8$ for QS and SF galaxies, respectively. This is consistent with studies in literature \cite{Davidzon2013,Siudek2017} that found that $\mathrm{\log{(M_{*}/M_{\odot})}} \sim 10.3$ at $\mathrm{z \sim 0.3}$ is the transition mass that separates the SF galaxies (below the transition mass) and the QS galaxies (above transition mass) dominant stellar mass contribution. Furthermore, the e-folding time distributions are in agreement with the typical time-scales found in literature for QS galaxies ($0.1-0.3$ Gyr) and for SF galaxies ($1 - 5$ Gyr) \cite{Renzini2006,Schawinski2014,Siudek2017}.

We find consistent results between observations and simulations, independently of the galaxy type, if we take into account the median values and the median errors in table \ref{table:stellar_pop_values} for the QS and SF population. However, as it is clear from figures \ref{fig:tortorelli_fig11} and \ref{fig:tortorelli_fig12}, there are slight discrepancies in the distributions of certain parameters, both for QS and SF galaxies, that we investigate further. 

The SFR and the stellar mass distributions of QS and SF galaxies are in good agreement between observations and simulations, spanning a similar range in values. The distributions of the e-folding time for observed and simulated QS galaxies are slightly displaced one respect to another (see figure \ref{fig:tortorelli_fig11}), with simulations predicting a larger median value of $\tau$. On the contrary, for SF galaxies (see figure \ref{fig:tortorelli_fig12}), the median value of $\tau$ is larger for the observations. From figure \ref{fig:tortorelli_fig12}, we see that the smaller median $\tau$ value for simulations is due to an excess of low $\tau$ objects.

The distributions of attenuation and mass-weighted stellar age show a more marked disagreement for QS galaxies (see figure \ref{fig:tortorelli_fig11}). The simulations produce QS galaxies with lower attenuation and older ages with respect to observed QS galaxies. This might be due to the existing degeneracy between dust and age: dusty galaxies present red colours that lead the SED fitting codes to predict older ages. This is a well known problem in stellar population models (e.g., \cite{Takagi1999,Cidfernandes2005}). For SF galaxies, instead, the distributions of attenuation are marginally consistent between simulated and observed galaxies. However, for SF galaxies, we see discrepancies in the distributions of ages (similarly as for QS), where the effect of having older ages predicted by the simulations is mainly due to an excess of simulated galaxies with low $\tau$. Despite these discrepancies might be attributed to the degeneracy among dust, metallicity and ages \cite{Worthey1999}, it is well know that the SED fitting has larger errors on these parameter estimates with respect to spectroscopy (see e.g., \cite{Johnson2021}).

The results shown in this section proves that our simulated galaxies are able to correctly reproduce the observed bimodality in properties between QS and SF galaxies. However, some discrepancies are present in the attenuation, mass-weighted stellar age and e-folding time. In the next section, we check whether we can improve in these parameter estimates by performing an SED fitting where we let the metallicity free to vary.

\begin{figure}[t!]
\centering
\includegraphics[width=12cm]{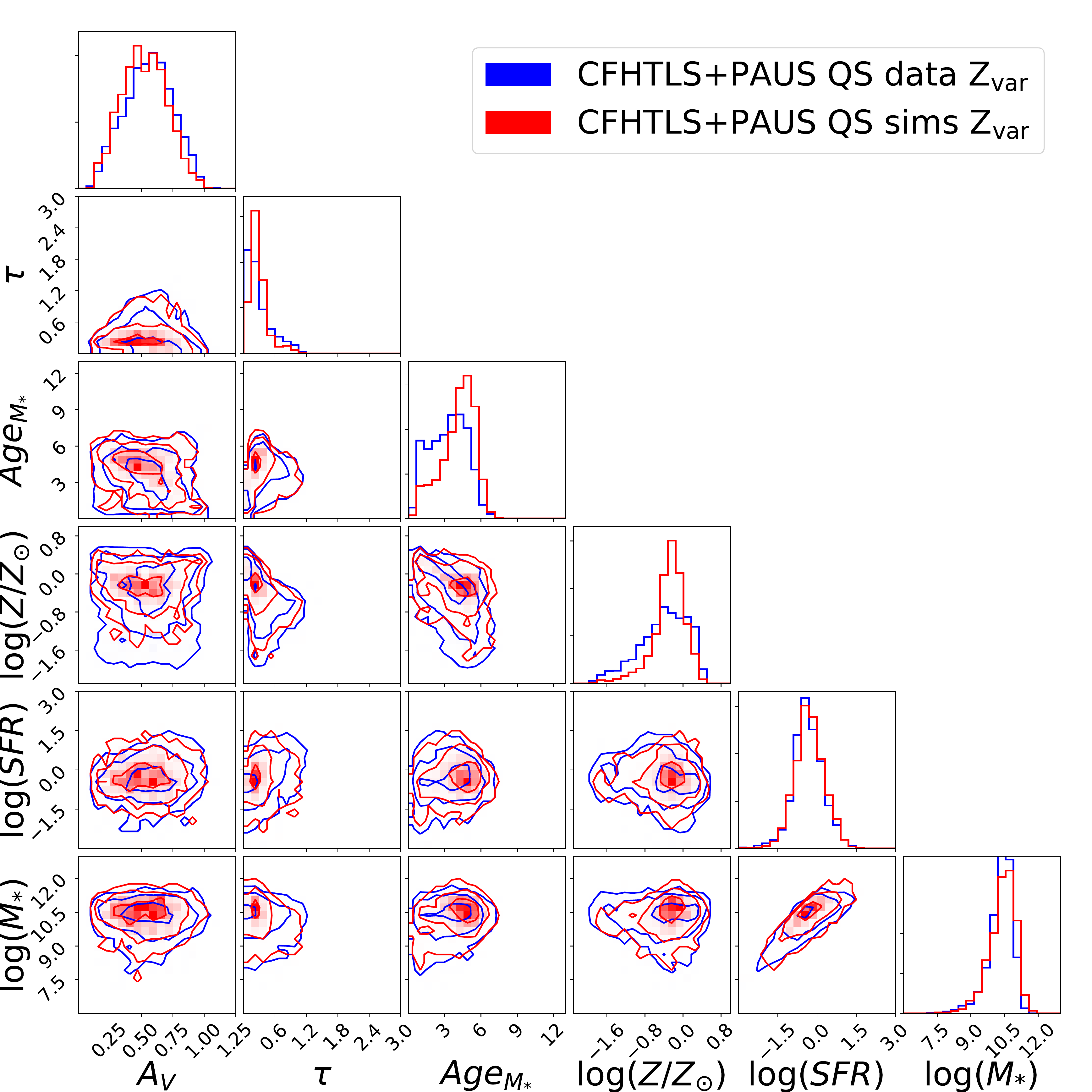}
\caption{This figure shows the comparison between the stellar population properties of QS observed (blue) and simulated (red) galaxies. The properties are measured using the SED fitting code \textsc{CIGALE} with variable metallicity. The fit uses the broad-band photometric information from CFHTLS and narrow-band photometric information from PAUS. $A_V$ is in units of mag, $\tau$ and $\mathrm{Age_{M_{*}}}$ in units of Gyr, SFR in units of $\mathrm{M_{\odot} / yr}$, while $\mathrm{\log{(M_{*})}}$ is in units of $\mathrm{M_{\odot}}$. $\mathrm{\log{(Z/Z_{\odot})}}$ is the metallicity expressed in solar units.}
\label{fig:tortorelli_fig13}
\end{figure}

\begin{figure}[t!]
\centering
\includegraphics[width=12cm]{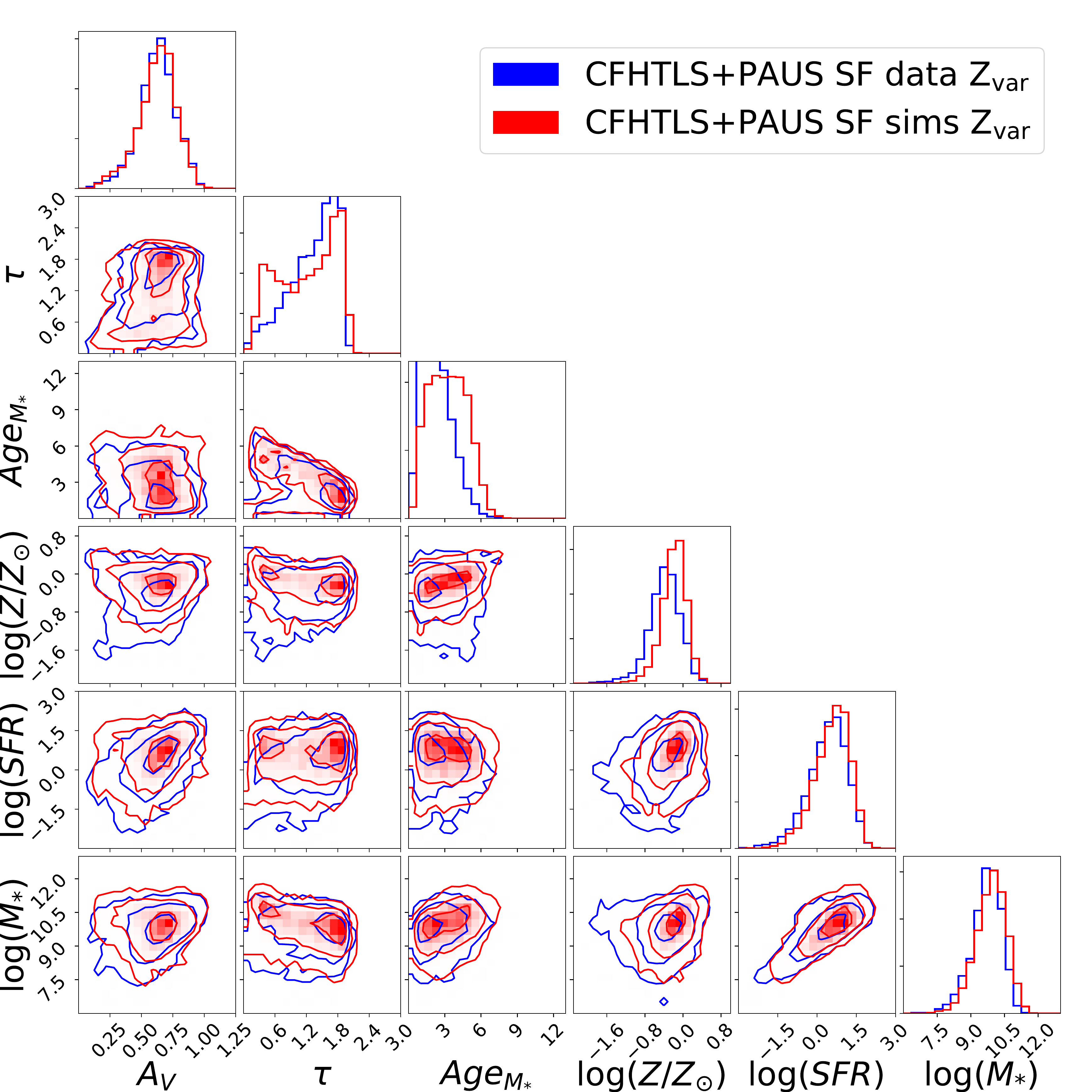}
\caption{This plot shows the same comparison as in figure \ref{fig:tortorelli_fig13}, but for SF galaxies.}
\label{fig:tortorelli_fig14}
\end{figure}

\subsection{SED fitting with variable metallicity}
\label{section:sed_variable}

Figures \ref{fig:tortorelli_fig13} and \ref{fig:tortorelli_fig14} show the results of performing an SED fitting with \textsc{CIGALE} on broad-bands and narrow-bands where we consider the metallicity as a free parameter. We measure the same stellar population properties as in section \ref{section:sed_solar}, with the addition of the metallicity expressed in units of solar metallicity, $\mathrm{\log{(Z/Z_{\odot})}}$.  The median values and the median errors are reported in table \ref{table:stellar_pop_values}. Also in this case, by looking at the median values and median errors alone, all the parameters are consistent between observations and simulations. 

Despite having larger median errors, the stellar mass and the SFR distributions for QS galaxies are in a better agreement than those obtained with the SED fit with solar metallicity. This is confirmed by the smaller absolute difference in the median values between observations and simulations with respect to the previous estimates. Both the observed and the simulated QS galaxies are slightly more massive and with higher SFRs (both by less than$\sim 0.1$ dex) than those described in the previous section. The stellar mass distributions of SF galaxies are, instead, almost unchanged. The same considerations about the transition mass highlighted in section \ref{section:sed_solar} apply here \cite{Davidzon2013,Siudek2017}. Interestingly, the SFR of the observed SF galaxies has a median value that is smaller than the simulations and smaller than the estimate based on solar metallicity. However, we keep finding that the SF galaxies are less massive and with higher SFR than QS galaxies, as expected from literature studies.

The e-folding times of QS galaxies have larger median values with respect to their estimates based on solar metallicity, but their absolute difference is smaller. The e-folding time distribution of SF simulated galaxies is still showing an excess of objects with low $\tau$. This makes the median value of $\tau$ for SF simulated galaxies to be smaller than the corresponding value for SF observed galaxies. Furthermore, the values for SF galaxies are smaller than those estimated in section \ref{section:sed_solar}. However, the median values for QS and SF galaxies are still consistent with the studies in literature \cite{Renzini2006,Siudek2017}.

By letting the metallicity free to vary, the attenuation distributions of QS and SF galaxies are in very good agreement between observations and simulations. The effect of the varying metallicity is also visible on the age estimate. Both the QS and the SF observed and simulated galaxies are older than their estimates in section \ref{section:sed_solar}. Since the attenuations are consistent, the age-metallicity degeneracy favours models that have sub-solar metallicity. QS observed and simulated galaxies are roughly $\sim 1$ Gyr older than SF galaxies, but the metallicities span a similar range of values. At fixed galaxy type, simulated galaxies are older and more metal rich than observed galaxies. The addition of the free metallicity in the SED fit fixes the discrepancy in the attenuations, but not the discrepancy in the mass-weighted stellar ages and in the contamination of low $\tau$ objects.

Although mass-weighted stellar ages and metallicities show distributions which seem discrepant, the large error on these parameter estimates makes the parameters still consistent within errors. The large errors on mass-weighted stellar ages and metallicities estimated with SED fitting is known in literature \cite{Johnson2021}. The SED fitting procedure is robust in capturing the amplitude of the galaxy SEDs. Therefore, we expect that it performs well in constraining global properties, such as stellar masses and SFRs, while it lacks in resolving power to constrain the spectral features that are necessary to estimate the mass-weighted stellar ages and metallicities. We show in Appendix \ref{appendix:sed_constraining_power} that how well the parameters are determined from SED fitting is the main driver of the lack of constraining power on ages, dust and metallicities compared to their degeneracies.

The results show that the simulated galaxies drawn from the calibrated galaxy population model exhibit the observed bimodality of QS and SF galaxies physical properties. QS galaxies are older, more massive, less star-forming and less attenuated by dust than SF galaxies. This shows the quality of our galaxy population model that is able to output a population of objects which have a similar distribution of physical properties than the observed galaxies, even though the physical properties are not directly drawn from the model itself.

\section{Conclusions}
\label{section:conclusions}

Narrow-band galaxy surveys are a promising way ahead for future cosmology and galaxy evolution studies. They sit in between broad-band photometry and spectroscopy and they offer several advantages with respect to those, e.g., better spectral resolution than broad-bands and lack of target selection. With the increasing amount of data and complexity that these narrow-band surveys provide, new methods to robustly measure galaxy population properties need to be developed.

In this paper, we use data from the Physics of the Accelerated Universe Survey (PAUS) to constrain the coefficients used to create the galaxy spectral energy distributions of the empirical galaxy population model presented in \cite{herbel17,Tortorelli2020}. In \cite{Tortorelli2020}, we forward-modelled the Canada-France-Hawaii Telescope Legacy Survey (CFHTLS) and we constrained the luminosity function and size parameters of the model using Approximate Bayesian computation (ABC), given that the likelihood was not empirically tractable. The broad-band dataset left however the spectral part of the galaxy population model only broadly constrained. PAUS spectral resolution constitutes the key element to pin down that part. 

We constrain the full empirical galaxy population model using ABC applied to CFHTLS and PAUS images. We use as prior for this work a resampled version of the posterior distribution presented in \cite{Tortorelli2020}. We use an iterative approach where at each iteration we perform a Rejection ABC algorithm to constrain the model parameters. For each set of parameters, we simulate CFHTLS and PAUS images, we match the catalogues from these two and measure different distance metrics. By thresholding the distance metrics based on the 10-th percentile value, we obtain the approximate Bayesian posterior for all the parameters of the galaxy population model, both for blue and red galaxies. The luminosity function and size parameters do not gain much in terms of constraints, since CFHTLS data already saturated the available information in the ABC run in \cite{Tortorelli2020}. PAUS data, instead, are able to provide the additional spectral resolution needed to constrain the spectral coefficients.

To test the obtained approximate Bayesian posterior, we perform a Principal Component Analysis (PCA) on the observed-frame and rest-frame observed and simulated narrow-band spectra. The PCA allows us to capture the global spectral information from the PAUS dataset and it is the preferred approach since the large number of wavelength bins makes the use of colours not an efficient option. We find the Principal Components (PCs) of observed data and simulations and we compare them through the `mixing matrix'. The latter is the matrix originated from the scalar product between the observed and simulated PCs. For the observed-frame analysis, the median spectra of observed and simulated galaxies and the first component of the mixing matrix are in very good agreement. Higher order PCs are instead noise dominated. This is due to the fact that the individual spectral features are smoothed by the different redshifts at which they appear when performing the PCA in the observed-frame. The constraint is therefore on the overall agreement of the median values and dispersion of the observed-frame spectral shapes. By performing the same analysis in the rest-frame, the spectral features fall in the same wavelength bin and their signal sums up. The resulting mixing matrix shows a remarkably good agreement in all the 5 PCs and in the median spectra. This confirms that the constraints we obtain on the galaxy population model with CFHTLS and PAUS data allows us to create simulated galaxies that are characterised by realistic spectral features.

The large object statistic and the spectral resolution of narrow-band data makes them particularly interesting to study the galaxy stellar population properties. There are still very few attempts in the literature to measure galaxy properties with spectral energy distribution (SED) fitting on narrow-band data and this work constitutes one of the first steps towards this direction. We use observed galaxies and simulated galaxies from the approximate Bayesian posterior with magnitude $18 \le i^{'} \le 22.5$ for this analysis. We perform SED fitting on them using \textsc{CIGALE}. We fit galaxies using the CFHTLS broad-bands and the PAUS narrow-bands, both by fixing the metallicity to the solar value and by letting it free to vary. With SED fitting, we further test the quality of our simulated narrow-band spectra and the ability to predict galaxy physical properties that are not directly drawn from the model.

The results show that the simulated galaxies show the bimodality of quiescent and star-forming galaxies that is seen in other studies in literature. Quiescent galaxies appear to be older, more massive, with lower star-formation rates and less attenuated by dust than star-forming galaxies. The stellar masses and the star-formation rates are in very good agreement between observations and simulations, for both quiescent and star-forming galaxies. The mass-weighted stellar ages and metallicities show some discrepancies in the distributions, although they are consistent within errors. The SED fitting provides large errors on the mass-weighted stellar age and metallicity estimates, while it is more robust in the stellar mass and star-formation rate estimates. This is not surprising, since the SED fitting is more accurate in capturing the amplitude of the galaxy spectral shapes, rather than measuring the individual spectral features, which are those that constrain the stellar ages and metallicities.

Even though the empirical galaxy population model contains simple yet realistic prescription for galaxy properties, the combination of model plus image simulation is able to correctly forward-model very diverse and information-rich dataset such as PAUS and CFHTLS. Furthermore, it is able to correctly reproduce the diversity of galaxy properties that can be probed with these surveys. Using PAUS and CFHTLS, we are able to provide a fully calibrated galaxy population model from which we can draw realistic simulated galaxies for cosmology and galaxy evolution studies. These galaxies constitute a realistic representation of the galaxy population at the redshifts spanned by the data employed. In the spirit of forward-modelling, the galaxy population model can be further extended and made more complex, if needed. Progress in this direction will be made with the implementation of a realistic clustering prescription and more complex stellar population templates that might be required when studying the higher redshift galaxy population.

\acknowledgments

We acknowledge support by Swiss National Science Foundation (SNF) grant 200021\_169130. LT is grateful to Dominik Z\"urcher and Uwe Schmitt for useful discussions. The PAU Survey is partially supported by MINECO under grants CSD2007-00060, AYA2015-71825, ESP2017-89838, PGC2018-094773, PGC2018-102021, PID2019-111317GB, SEV-2016-0588, SEV-2016-0597, MDM-2015-0509 and Juan de la Cierva fellowship and LACEGAL and EWC Marie Sklodowska-Curie grant No 734374 and no.776247 with ERDF funds from the EU Horizon 2020 Programme, some of which include ERDF funds from the European Union. IEEC and IFAE are partially funded by the CERCA and Beatriu de Pinos program of the Generalitat de Catalunya. Funding for PAUS has also been provided by Durham University (via the ERC StG DEGAS-259586), ETH Zurich, Leiden University (via ERC StG ADULT-279396 and Netherlands Organisation for Scientific Research (NWO) Vici grant 639.043.512), University College London and from the European Union's Horizon 2020 research and innovation programme under the grant agreement No 776247 EWC. The PAU data center is hosted by the Port d'Informaci\'o Cient\'ifica (PIC), maintained through a collaboration of CIEMAT and IFAE, with additional support from Universitat Aut\'onoma de Barcelona and ERDF. We acknowledge the PIC services department team for their support and fruitful discussions. M. Siudek's work has been supported by the European Union's Horizon 2020 Research and Innovation Programme under the Maria Sklodowska-Curie grant agreement (No. 754510), the Polish National Science Centre (UMO-2016/23/N/ST9/02963), the Spanish Ministry of Science and Innovation through the Juan de la Cierva-formacion programme (FJC2018-038792-I). H. Hildebrandt is supported by a Heisenberg grant of the Deutsche Forschungsgemeinschaft (Hi 1495/5-1) as well as an ERC Consolidator Grant (No. 770935). A. Amara is supported by a Royal Society Wolfson Fellowship. A.H. Wright is supported by an ERC Consolidator Grant (No. 770935). Based on observations obtained with MegaPrime/MegaCam, a joint project of CFHT and CEA/IRFU, at the Canada-France-Hawaii Telescope (CFHT) which is operated by the National Research Council (NRC) of Canada, the Institut National des Science de l'Univers of the Centre National de la Recherche Scientifique (CNRS) of France, and the University of Hawaii. This work is based in part on data products produced at Terapix available at the Canadian Astronomy Data Centre as part of the Canada-France-Hawaii Telescope Legacy Survey, a collaborative project of NRC and CNRS. This work has made use of data from the European Space Agency (ESA) mission {\it Gaia} (\url{https://www.cosmos.esa.int/gaia}), processed by the {\it Gaia} Data Processing and Analysis Consortium (DPAC, \url{https://www.cosmos.esa.int/web/gaia/dpac/consortium}). Funding for the DPAC has been provided by national institutions, in particular the institutions participating in the {\it Gaia} Multilateral Agreement. This work has made use of CosmoHub. CosmoHub has been developed by the Port d'Informaci\'o Cient\'ifica (PIC), maintained through a collaboration of the Institut de F\'isica d'Altes Energies (IFAE) and the Centro de Investigaciones Energ\'eticas, Medioambientales y Tecnol\'ogicas (CIEMAT) and the Institute of Space Sciences (CSIC \& IEEC), and was partially funded by the "Plan Estatal de Investigaci\'on Cient\'ifica y T\'ecnica y de Innovaci\'on" program of the Spanish government.

\newpage

\appendix

\section{Quiescent and Star-forming galaxies separation}
\label{appendix:uvj_selection}

\begin{figure}[t!]
\centering
\includegraphics[width=14cm]{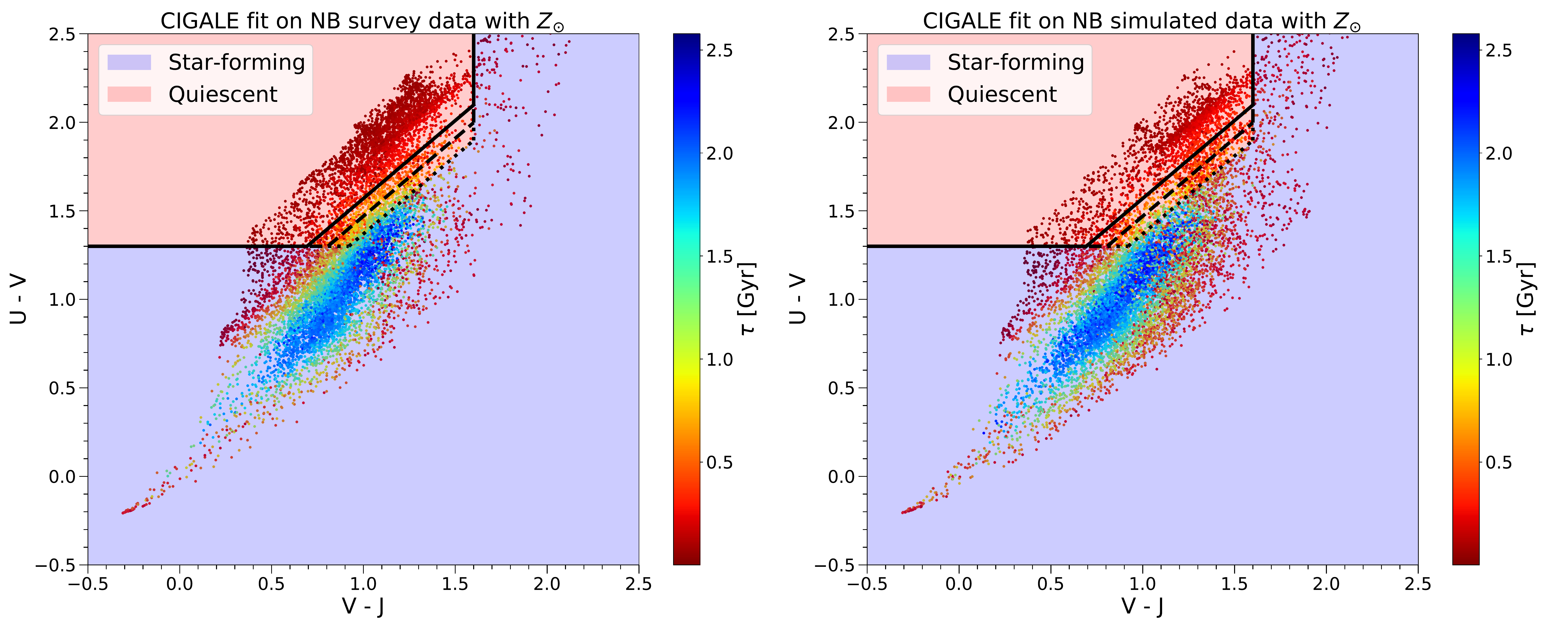}
\includegraphics[width=14cm]{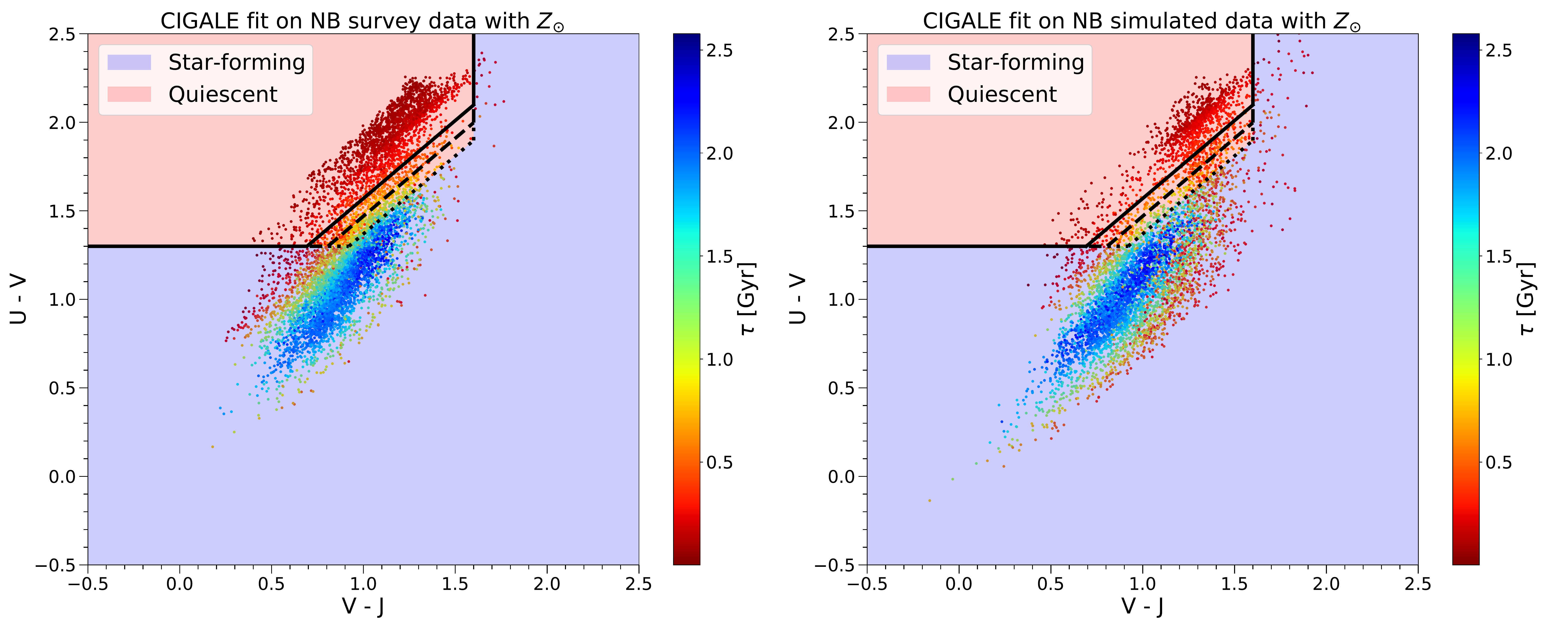}
\caption{This figure shows the UVJ diagram obtained from the rest-frame U, V and J magnitudes from the SED fit on broad-bands and narrow-bands with solar metallicity. The left panels show the UVJ diagram for observed galaxies, while the right panels for simulated ones. Upper panels show all galaxies, while bottom panels show galaxies with reduced chi-square,  $\chi^2_{\mathrm{r}} < \bar{\chi^2_{\mathrm{r}}} + 3\sigma$. Galaxies are colour-coded according to their $\tau$ values, with high $\tau$ objects (blue points) being star-forming and low $\tau$ objects (red points) quiescent. Lines separate the UVJ diagram into the star-forming (light blue) and quiescent (light red) regions according to the criteria in \cite{Williams2009}. The solid, dashed and dotted lines refer to the criteria in the redshift ranges $\mathrm{0.0 < z < 0.5}$, $\mathrm{0.5 < z < 1.0}$ and $\mathrm{1.0 < z < 2.0}$, respectively.}
\label{fig:tortorelli_fig15}
\end{figure}

In our empirical galaxy population model, we draw red and blue galaxies from two different redshift-dependent luminosity functions. In appendix D of \cite{Tortorelli2020} we showed that this definition of red and blue galaxies is consistent with the separation in quiescent (QS) and star-forming (SF) objects via the rest-frame V - J vs U - V colour space (UVJ diagram, \cite{Wuyts2007,Williams2009}). Therefore, in this work, we use the UVJ diagram to separate both observed and simulated galaxies into SF and QS objects. The rest-frame U - V and V - J colours are obtained from the SED fit with \textsc{CIGALE} by integrating the best-fitting spectra in the U, V and J Johnson photometric system bands.

Figure \ref{fig:tortorelli_fig15} shows the UVJ diagram for observed (left panels) and simulated (right panels) galaxies obtained from the SED fitting of CFHTLS broad-bands and PAUS narrow-bands with solar metallicity. Galaxies are colour-coded according to the e-folding time $\tau$ of the delayed part of the SFH. High $\tau$ objects experience a prolonged process of star-formation that is still active nowadays, while low $\tau$ objects are passively evolving. The black lines separate the UVJ diagram into a region occupied by SF galaxies (light blue regions) and another occupied by QS galaxies (light red regions). The reported selection criteria is based on \cite{Williams2009}. The horizontal $\mathrm{U - V > 1.3}$ and vertical $\mathrm{V - J < 1.6}$ lines are redshift-independent and prevent the contamination of the sample from unobscured and dusty SFGs, respectively. The diagonal lines, instead, depend on redshift
\begin{equation}
\begin{split}
\mathrm{(U - V)} >& \ 0.88 \times \mathrm{(V - J)} + 0.69, \quad [0.0 < z < 0.5]\\ %solid
\mathrm{(U - V)} >& \ 0.88 \times \mathrm{(V - J)} + 0.59, \quad [0.5 < z < 1.0]\\ % dashed
\mathrm{(U - V)} >& \ 0.88 \times \mathrm{(V - J)} + 0.49, \quad [1.0 < z < 2.0]\\ % dotted
\end{split}
\end{equation}
The solid, dashed and dotted black lines refer to the separation in the redshift ranges $\mathrm{0.0 < z < 0.5}$, $\mathrm{0.5 < z < 1.0}$ and $\mathrm{1.0 < z < 2.0}$, respectively. 

\begin{figure}[t!]
\centering
\includegraphics[width=14cm]{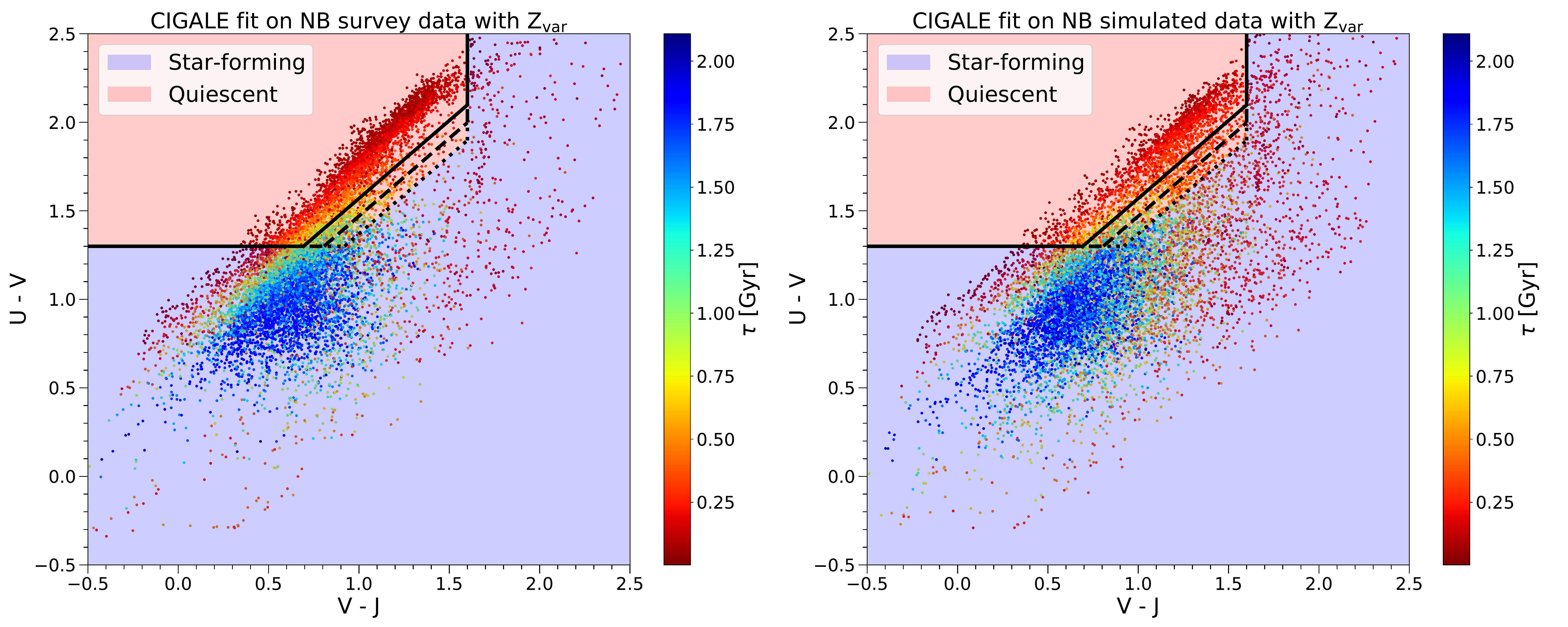}
\includegraphics[width=14cm]{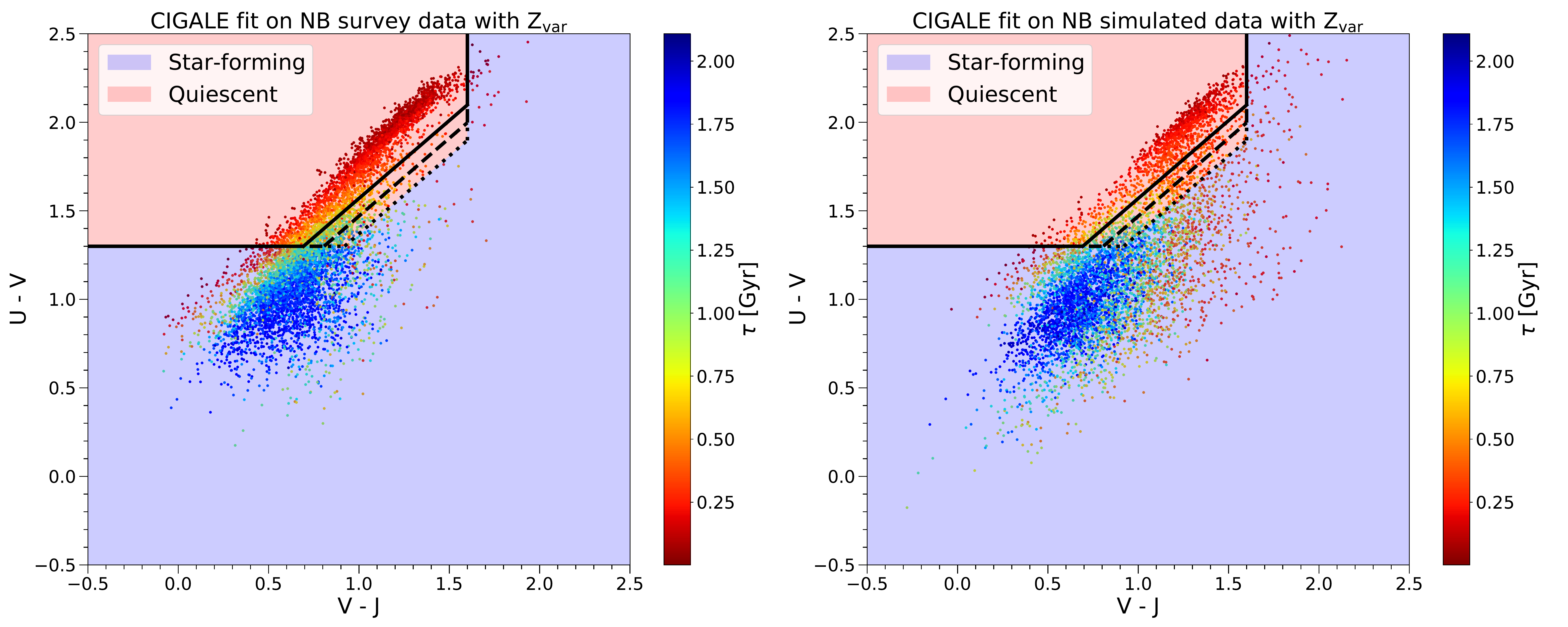}
\caption{This figure shows the UVJ diagram obtained from the rest-frame U, V and J magnitudes from the SED fit on broad-bands and narrow-bands with variable metallicity. The left panels show the UVJ diagram for observed galaxies, while the right panels for simulated ones. Upper panels show all galaxies, while bottom panels show galaxies with reduced chi-square, $\chi^2_{\mathrm{r}} < \bar{\chi^2_{\mathrm{r}}} + 3\sigma$. Galaxies are colour-coded according to their $\tau$ values, with high $\tau$ objects (blue points) being star-forming and low $\tau$ objects (red points) quiescent. Lines separate the UVJ diagram into the star-forming (light blue) and quiescent (light red) regions according to the criteria in \cite{Williams2009}. The solid, dashed and dotted lines refer to the criteria in the redshift ranges $\mathrm{0.0 < z < 0.5}$, $\mathrm{0.5 < z < 1.0}$ and $\mathrm{1.0 < z < 2.0}$, respectively.}
\label{fig:tortorelli_fig16}
\end{figure}

The bulk of the high $\tau$ (blue points) and the low $\tau$ (red points) objects correctly overlaps with the regions where SF and QS galaxies are expected to be according to the UVJ separation. However, in the upper panels of figure \ref{fig:tortorelli_fig15}, we see a contamination of low $\tau$ objects in the region that should be occupied by SF galaxies. This happens both for observed and simulated galaxies. 

We reduce this contamination by selecting the objects with a good global quality of the fitted SEDs, expressed via the reduced chi-square $\chi^2_{\mathrm{r}}$. We remove outliers from the $\chi^2_{\mathrm{r}}$ value distributions of observed and simulated galaxies and keep only those objects for which $\chi^2_{\mathrm{r}} < \bar{\chi^2_{\mathrm{r}}} + 3\sigma$ \cite{Malek2014,Malek2017,Malek2018}, where $\bar{\chi^2_{\mathrm{r}}}$ is the mean reduced chi-square and $\sigma$ is its standard deviation. The lower panels of figure \ref{fig:tortorelli_fig15} shows that we remove the majority of the contamination for both observations and simulations. However, there is some residual contamination, especially for the simulations, that causes the low $\tau$ tail discussed in figures \ref{fig:tortorelli_fig12} and \ref{fig:tortorelli_fig14} of sections \ref{section:sed_solar} and \ref{section:sed_variable}. We try to further restrict the $\chi^2_{\mathrm{r}}$ selection, but the contamination still persists. Therefore, we keep the selection introduced earlier since it has already been adopted in literature \cite{Malek2014,Malek2017,Malek2018}.

The contamination visible in figure \ref{fig:tortorelli_fig15} may be due to the different attenuations observations and simulations show in figures \ref{fig:tortorelli_fig11} and \ref{fig:tortorelli_fig12} when performing the SED fitting with solar metallicity. Therefore, we investigate whether this contamination still holds when we let the metallicity free to vary, since in this case the attenuations are in agreement. 

The UVJ diagram obtained from the SED fitting with variable metallicity is shown in figure \ref{fig:tortorelli_fig16}. The upper panels show the distribution of all the fitted objects for observations (upper left panel) and simulations (upper right panel), while the lower panels show the distribution for the reduced chi-square selected galaxies. The bulk of the high $\tau$ (blue points) and the low $\tau$ (red points) galaxies is in agreement with the regions separating SF and QS objects. However, the contamination of low $\tau$ galaxies in the region of SF objects still persist. 

Both in the case of solar and variable metallicity, these contaminating objects have a good value of $\chi^2_{\mathrm{r}}$. Therefore, the reason for their position in the UVJ diagram is mainly due to the intrinsic degeneracy of the SED fitting results when we lack photometric points corresponding to the rest-frame J band.

\section{SED fitting constraining power}
\label{appendix:sed_constraining_power}

\begin{figure}[t!]
\centering
\includegraphics[width=5cm]{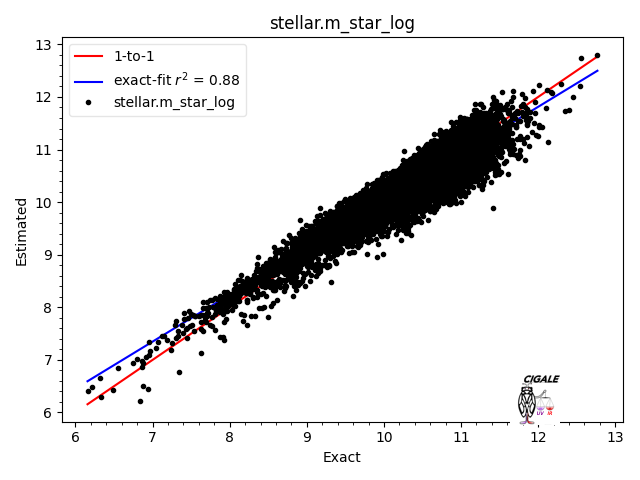}
\includegraphics[width=5cm]{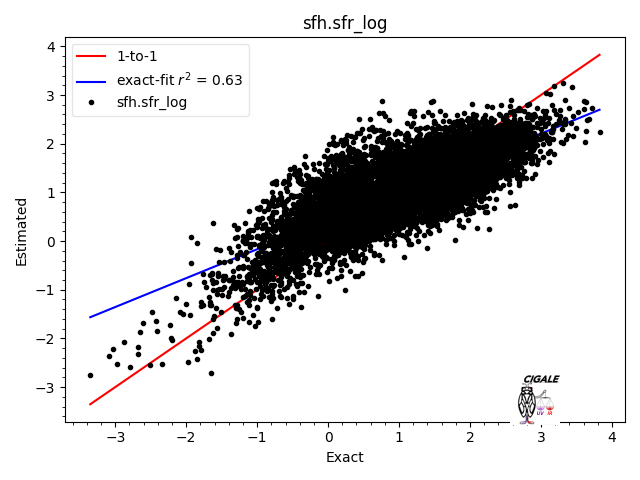}
\includegraphics[width=5cm]{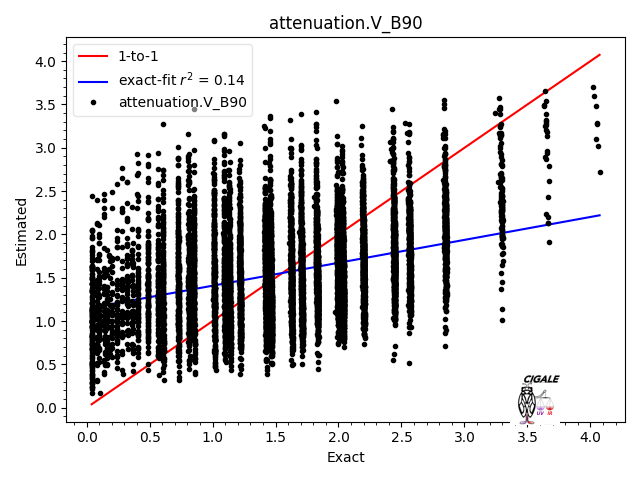}
\includegraphics[width=5cm]{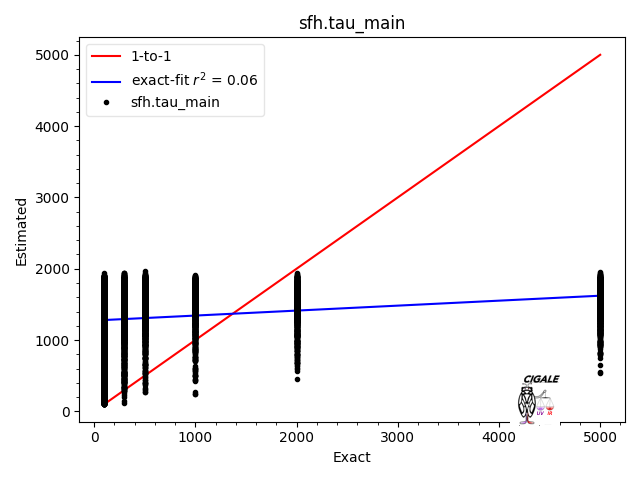}
\includegraphics[width=5cm]{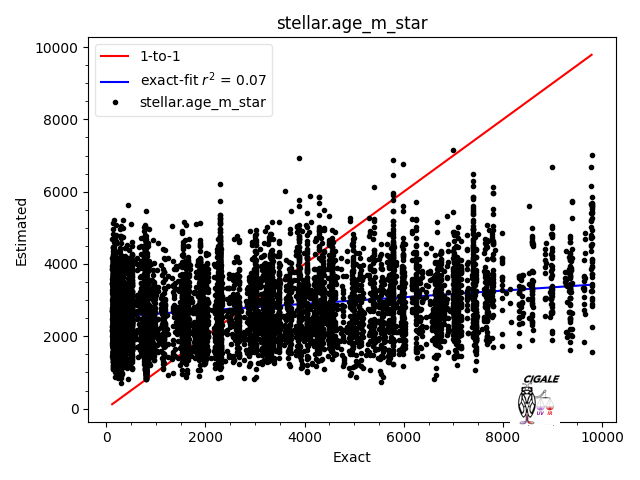}
\includegraphics[width=5cm]{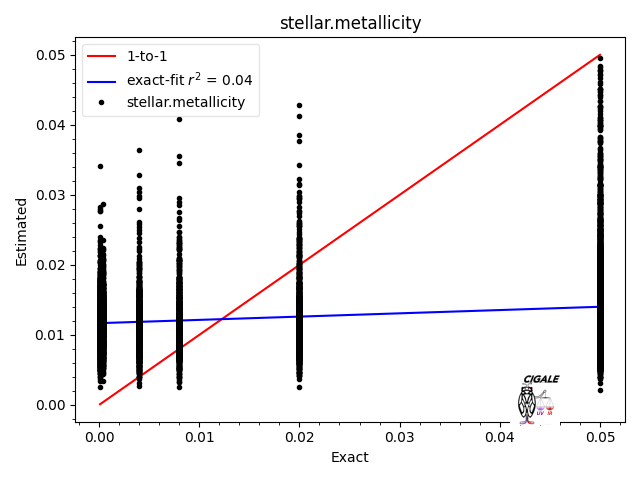}
\caption{This figure shows the ability of \textsc{CIGALE} in recovering the stellar population parameters. The horizontal axes show the true values of the parameters drawn from a Gaussian distribution centred on the parameters and with standard deviations equal to the uncertainty on the observations. The vertical axes is its \textsc{CIGALE} estimate. The red line is the one-to-one line of perfect recovery. The blue line is the linear fit to the recovered parameters. From top left to bottom right, this is shown for the stellar mass, SFR, dust content, e-folding time, age and metallicity. This is an example for the SED fitting performed on data with variable metallicity, but the same conclusions apply also to simulations.}
\label{fig:tortorelli_fig17}
\end{figure}

The SED fitting is able to correctly constrain the stellar mass and the SFR of a galaxy. However, due to the lack of wavelength resolution of photometric data, the spectral features necessary to estimate the stellar age and metallicity of a galaxy are less constrained.

This is confirmed by the diagnostic checks that \textsc{CIGALE} internally produces after each run \cite{Boquien2019}. In those checks, \textsc{CIGALE} tries to recover the true value of the best-fitting parameters by adding a value taken from a Gaussian distribution with the same standard deviation as the uncertainty on the observations and analysed in the same way.

As shown in figure \ref{fig:tortorelli_fig17}, CIGALE recovers with high accuracy the SFRs and stellar masses of galaxies, with marginally good accuracy dust and with low accuracy e-folding times, ages and metallicities, independently of whether we consider data or simulations. This implies that the main driver of the lack of constraining power on ages and metallicities for data and simulations is how well the SED fitting with CIGALE is actually able to constrain them. The optimisation of CIGALE and the testing of other SED fitting codes for the estimates of stellar population properties with narrow-bands and connecting them to more detailed galaxy classification \cite{Siudek2018a,Siudek2018b} are the subjects of a future work.


\bibliographystyle{unsrt}
\bibliography{tortorelli_paus_bibliography}

% Please avoid comments such as "For a review'', "For some examples",
% "and references therein" or move them in the text. In general,
% please leave only references in the bibliography and move all
% accessory text in footnotes.

% Also, please have only one work for each \bibitem.

\end{document}